\documentclass[aps,reprint,superscriptaddress,nofootinbib,showpacs,twocolumn]{revtex4-1}
\usepackage{amsmath,latexsym,amssymb,hyperref,graphicx, slashed,color}

\usepackage{mathrsfs}

\usepackage[T1]{fontenc}

\hypersetup{colorlinks, citecolor=nicegreen,linkcolor=nicered, urlcolor=darkblue}
\definecolor{nicered}{rgb}{.7,.1,.1}
\definecolor{nicegreen}{rgb}{.2,.7,.1}
\definecolor{lightgreen}{rgb}{.6,.9,.5}
\definecolor{darkblue}{rgb}{0,0,.5}
\definecolor{el}{rgb}{.9, .8, .7}
\definecolor{mu}{rgb}{.8, .7, .8}

\definecolor{red}{rgb}{1,0,0}

\newcommand\GeV{\text{GeV}}
\newcommand\TeV{\text{TeV}}
\newcommand\meV{\text{meV}}

\newcommand\SEC[1]{\medskip\noindent{\sl\bfseries #1}}

\newcommand{\be}{\begin{equation}}
\newcommand{\ee}{\end{equation}}
\newcommand{\bea}{\begin{eqnarray}}
\newcommand{\eea}{\end{eqnarray}}

\newcommand\s{\sigma}

\newcommand\C{\mathcal C}
\renewcommand\P{\mathcal P}

\renewcommand{\b}{\beta}
\renewcommand{\t}{\theta}

\begin{document}

\title{Left-Right Symmetry at FCC-hh}

\author{Miha Nemev\v{s}ek}
\email{miha.nemevsek@ijs.si}
\affiliation{\normalsize \it  Jo\v zef Stefan Institute, Jamova 39, 1000 Ljubljana, Slovenia}
\affiliation{\normalsize \it  Faculty of Mathematics and Physics, University of Ljubljana, Jadranska 19, 1000 Ljubljana, Slovenia}

\author{Fabrizio Nesti}
\email{fabrizio.nesti@aquila.infn.it}
\affiliation{\normalsize \it 
  Dipartimento di Scienze Fisiche e Chimiche, Universit\`a dell'Aquila, via Vetoio, I-67100, L'Aquila, Italy}
\affiliation{\normalsize \it INFN, Laboratori Nazionali del Gran Sasso, I-67100 Assergi (AQ), Italy}

\date{\today}

\vspace{1cm}

\begin{abstract} \noindent 
  We study the production of right-handed $W_R$ bosons and heavy neutrinos $N$ at a future 100\,TeV high energy 
  hadron collider in the context of Left-Right symmetry, including the effects of $W_L-W_R$ gauge-boson mixing.
  We estimate the collider reach for up to 3/ab integrated luminosity using a multi-binned sensitivity measure.
  In the Keung-Senjanovi\'c and missing energy channels, the 3$\,\s$
  sensitivity extends up to $M_{W_R}=35$ and 37\,TeV, respectively.
  We further clarify the interplay between the missing energy channel and the (expected) limits from neutrinoless double
  beta decay searches, Big Bang nucleosynthesis (and dark matter).
\end{abstract}



\maketitle


\section{Introduction} \label{sec:Intro}
\noindent
With the enduring experimental successes of the Standard Model (SM), it is striking that we still
lack a definitive theory of neutrino masses.  
A hint for going beyond the SM might be found in its structure, where the fermion quantum numbers seem to 
point to an underlying parity symmetric theory.
This is in sharp contrast with the maximal breaking of parity observed in the weak sector.  
This clash was resolved in the Left-Right (LR) symmetric theories~\cite{Pati:1974yy, Mohapatra:1974hk,
Senjanovic:1975rk, Senjanovic:1978ev, Minkowski:1977sc, Mohapatra:1979ia} and turned out to
be deeply connected with the issue of neutrino mass origin.

In the minimal LR symmetric model (LRSM) parity is broken spontaneously~\cite{Senjanovic:1975rk, 
Senjanovic:1978ev, Minkowski:1977sc, Mohapatra:1979ia}, together with the new right-handed (RH) 
weak gauge group $SU(2)_R$.
The fermion sector then keeps the parity symmetry, while the gauge sector does not.
Spontaneous symmetry breaking is triggered by a $SU(2)_R$-triplet scalar $\Delta_R$ that simultaneously 
generates the masses of additional gauge bosons $W_R$ and $Z_{LR}$, as well as the masses for 
RH neutrinos $N$.
Their masses mainly come from a Majorana type Yukawa term that generates the $N$ mass and
breaks the total lepton number after $\Delta_R$ gets a vacuum expectation value (VEV).
The residual SM gauge group is then finally broken via a LR bi-doublet scalar field, which contains 
the SM Higgs doublet $h$ and an extra heavier doublet $H$.  
The bi-doublet has two VEVs that may give rise to a mixing of the SM $W$ and $W_R$.
After the completion of electroweak breaking, light neutrinos also get their Majorana masses with
contributions from the celebrated see-saw mechanism~\cite{Minkowski:1977sc, Mohapatra:1979ia,
Sawada:1979dis, Glashow:1979nm, Gell-Mann:1979vob}.
\medskip

\begin{figure}[b]
  \includegraphics[width=\columnwidth]{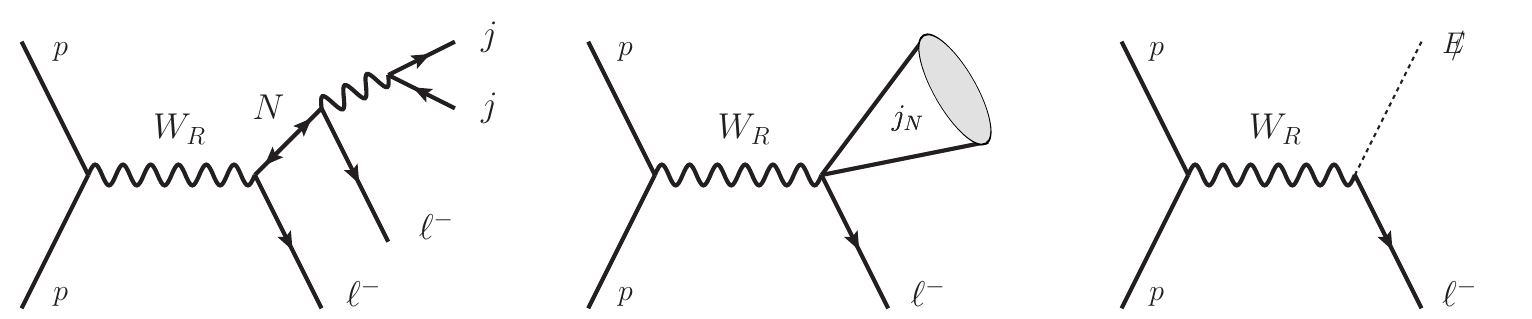}
  \caption{Final states of the $W_R$ production, with a prompt charged lepton plus $N$ decay products
    ranging from: lepton and jets (KS, left), the prompt merged $j_N$ (center) and the missing
    energy (right).}
\label{fig:layman}
\end{figure}

In general, to uncover the true microscopic picture of particle mass origin, we need to perform direct searches 
at colliders and measure the masses and couplings of elementary particles, just like we did with the Higgs boson.
Neutrinos are no exception and ultimately we would need to make a direct discovery at high energy colliders
to solidify our understanding of their mass origin.
Only such machines would allow us to perform direct searches for resonances, such as the $W_R$, and give
us immediate access to heavy Majorana neutrinos $N$.
In the golden Keung-Senjanovi\'c (KS) process~\cite{Keung:1983uu}, the $W_R$ is Drell-Yan produced 
and decays into a right-handed (RH) charged lepton $\ell_R$ and $N$, see~\cite{Cai:2017mow} for a review of LNV signals
at colliders.
In turn, $N$ decays dominantly through a possibly off-shell $W_R$ into another lepton and two
jets with the exact signal depending on its mass $m_N$~\cite{Ferrari:2000sp, Nemevsek:2011hz}, see
FIG.~\ref{fig:layman}.
Owing to the Majorana nature of $N$, the two leptons have the same electromagnetic charge half of the times, 
revealing the breaking of lepton number (see~\cite{Gluza:2016qqv} for departures from pure Majorana states). 
Lighter $N$ becomes boosted and its decay products collimate into a single neutrino jet~\cite{Mitra:2016kov}. 
Finally, if $N$ is below $\sim 100\,\GeV$ it becomes long lived and manifests itself as a lepton plus 
missing energy $\ell \slashed E$ signature~\cite{Nemevsek:2011hz}.  
The current LHC searches cover the range of well separated objects in $\ell \ell j j$~\cite{ATLAS:2018dcj,
CMS:2021dzb, ATLAS:2023cjo}, the collimated ``neutrino'' jets~\cite{CMS:2021dzb, ATLAS:2019isd,
ATLAS:2023cjo} or as a lepton with missing energy~\cite{CMS:2022krd}.

The above $W_R$ searches loose quickly sensitivity when $N$ is progressively off-shell, $m_N > M_{W_R}$.
In such case, one can resort to neutrino-less final states, such as di-jets and pairs of SM gauge bosons 
that appear in the presence of $W_L$--$W_R$ mixing.
The di-jet resonance searches were performed in~\cite{ATLAS:2019fgd, CMS:2019gwf}, the heavy
quark final state $W_R \to t b$ was looked for in~\cite{ATLAS:2021drn, CMS:2023ldh} and the 
$W_R \to W Z$ channel in~\cite{CMS:2022pjv}. 
All these limits converge into a current lower bound on $M_{W_R}$ in the range of $4$--$5.7 \text{ TeV}$.
The expected reach of the LHC can further extend to $6$--$7\,\TeV$ with large 
statistics~\cite{Nemevsek:2018bbt}, so the parameter space accessible by LHC is almost covered.
The aim of this work is to provide a definitive outlook for the 100 TeV hadronic colliders, connect
it to low energy processes and to the physics of the early universe.

\medskip

Apart from the existing collider searches, the precision frontier at low energies also delivers a set of
stringent constraints. 
It was known since the early days~\cite{Beall:1981ze} that loop-induced flavor changing processes in the 
$K$ meson sector push the $W_R$ scale into the few TeV regime.
Moreover, the LRSM contains an additional doublet $H$ with flavor off-diagonal couplings that mediate
flavor-changing processes even at tree level, which push the LR scale even higher~\cite{Senjanovic:1979cta}. 
A number of subsequent works addressed these issues~\cite{Ecker:1985vv,  Mohapatra:1983ae, Zhang:2007fn, 
Maiezza:2010ic, Bertolini:2012pu}.
Most recent updates~\cite{Bertolini:2014sua} uncovered the dominant role of $B$-meson oscillations
and set the limit in the ballpark of $M_{W_R}\gtrsim 8\,\TeV$ and $M_H\gtrsim 20\,\TeV$.
Even if such $M_{W_R}$ may still be marginally or indirectly probed by the LHC, the heavier $H$
implies that the model has to live at the brink of non-perturbativity~\cite{Maiezza:2016bzp, Maiezza:2016ybz}. 
A heavier $M_{W_R}$ would clearly relax this tension.

In parallel, constraints from CP violation come from the interplay between the neutron electric dipole
moment and meson processes~\cite{Maiezza:2014ala, Bertolini:2019out}.
These would require $M_{W_R}$ to be pushed beyond 10--20\,\TeV, at least if LR parity $\P$ is adopted,
see~\cite{Maiezza:2021dui} for the discussion of parity as gauge symmetry and the nature of its imposition.
And even if an axion is invoked, CP violation still implies lower bounds in the ballpark of
$10-20\,\TeV$~\cite{Bertolini:2020hjc}.
In case of $\C$ as LR parity, the additional CP phases are sufficient in order to accommodate
all the CP-violating channels and such bounds go away.

In summary, the LRSM scale is being driven to ever larger scales of $\mathcal O(10) \text{ TeV}$, 
nearly out of reach of the LHC, but easily probed by a future 100\,\TeV\ hadron collider (FCC).
A number of studies have started addressing this scenario~\cite{Rizzo:2014xma, Ng:2015hba,
Dev:2015kca, Mohapatra:2019qid}, see also~\cite{CidVidal:2018eel} and~\cite{Ruiz:2017nip}.
However, a complete assessment of the FCC potential for the LRSM, including the simulation of 
backgrounds and transitions between different regimes of $m_N$, is still missing.  
In this work we close this gap and clarify the FCC reach, by taking into account the standard KS 
and missing energy channels.

\medskip

In addition to the usual $W_R$ channels, we address the role of the LR gauge boson mixing
$\xi_{LR}$, that leads to an interplay between the production and decay via the SM $W$.
These channels are complemented by those mediated by Dirac Yukawa couplings that are responsible
for the mixing between the light and heavy Majorana neutrinos. 
With an input from neutrino oscillations and masses/mixings of $N$, one can disentangle the seesaw
and compute the Dirac mass matrix for both choices of LR parity: $\C$~\cite{Nemevsek:2012iq} and 
$\P$~\cite{Senjanovic:2016vxw, Senjanovic:2018xtu, Kiers:2022cyc}.
We show that their effect is relevant in the very light RH neutrino mass range.
Here, displaced signatures play a major role~\cite{Nemevsek:2018bbt} and shall be the subject of
dedicated studies, once the detector geometries and efficiencies are known for the FCC-hh.

Apart from to the involved analyses using displaced vertices, the missing energy signal can be understood
and estimated quite reliably.
It is precisely in this region of parameter space that interesting connections with other processes arise as well.
It turns out that the neutrinoless double beta ($0\nu\beta\beta$) decay rate from $N$ exchange~\cite{Tello:2010am}
and from additional mixed diagrams~\cite{Barry:2013xxa} is able to compete with FCC-hh, given the (optimistic) 
sensitivity of forthcoming experiments.
Finally, we should point out the connection to dark matter in the LRSM~\cite{Bezrukov:2009th, Nemevsek:2012cd} 
that may reside in the 20 TeV range~\cite{Nemevsek:2012cd} but is also subject
to additional constraints from large scale structures~\cite{Nemevsek:2022anh}.

\medskip

In Section~\ref{sec:Production} we review in detail the production of $W_R$ and the decay chains of the RH 
neutrino $N$.
In Section~\ref{sec:Backgrounds} we discuss the numerical simulations of relevant backgrounds.
In Section~\ref{sec:Signature} we analyze the signal features for the relevant processes, and in 
Section~\ref{sec:Sensitivity} we discuss the assessment of the expected sensitivity and present the results.  
Section~\ref{sec:Outlook} contains the final discussion and in the Appendix~\ref{app:DY} we give more details and 
analytic derivations.

%
%
\section{Production at the FCC and decay rates} \label{sec:Production}

\noindent
In this section we review the production of $W_R$ at a $pp$ collider, its decay through $N$, including
the various $N$ decay channels and the role of the left-right gauge boson mixing
\begin{equation} \label{eqWLWRMixing}
  \left| \xi_{LR} \right| \simeq \sin 2\b \left( \frac{M_W}{M_{W_R}} \right)^2 \, .
\end{equation}
Here, $t_\b\equiv \tan\b=v_2/v_1$ is the ratio of the two bi-doublet VEVs, see~\cite{Maiezza:2010ic} for
details.

%
\subsection{Production of $W_R$ \ldots} \label{sec:WR}

\noindent
The production of an on-shell $W_R$ proceeds through the Drell-Yan process involving the two 
initial partons
\begin{equation}
  \begin{split}
    \frac{\text {d}^2 \sigma_{p p \to W_R^+}}{\text{d} x_1 \text{d} x_2} &= 
    \frac{\pi^2 \alpha_2}{N_c}  \delta \left(\hat s - M_{W_R}^2 \right)\times
    \\
    & \sum_{u,d} |V_{ud}|^2 \left(f_u(x_1) f_{\overline d}(x_2) + 1 \leftrightarrow 2 \right) \, .
\end{split}
\end{equation}
Here, $x_{1,2}$ are the parton momentum fractions, $\hat s = x_1 x_2 s$ and $\sqrt s = 100 \text{ TeV}$ 
is the center of mass energy, see Appendix~\ref{app:DY} for a complete derivation.
             
The above formula also holds when the left-right gauge boson mixing $\xi_{LR}$ is turned on (via
$t_\b$) because the contributions from the right and the left-mixing currents, proportional to
$\sin^2 \xi_{LR}$ and $\cos^2 \xi_{LR}$ sum up to one, while the interference terms are suppressed
either by small quark masses or PDFs of the proton.

In Appendix~\ref{app:DY} we collect the rates of the various $W_R$ decay channels, namely dijet, $\ell N$, 
as well as $WZ/Wh$ mediated by gauge boson mixing.
We find that the parent $W_R$ is never produced with a high boost.
Indeed, we find that the maximal boost factor $\gamma$ is given by 
\begin{equation}
  \gamma_{W_R}^{\max} \simeq \frac{\sqrt s}{2 M_{W_R}} \, . 
\end{equation}
Moreover, the $W_R$ decay products are typically much lighter than $M_{W_R}$, such that they 
feature back-to-back geometry distinctive of two body decays.
The only relevant exception is the case of $\ell N$ with $N$ being nearly as heavy as $W_R$, 
to be discussed shortly below.
For completeness we report the $k$-factors for the production including NLO effects in 
Section~\ref{sec:Signature}.

%
%
\subsection{\ldots and $\ell N$} \label{sec:Signal}

\noindent
The triple differential cross section for the $pp \to \ell^+ N$ production via $W_R^+$ is given by
\begin{equation}
\begin{split} \label{eq:dsigdx12dth}
  \frac{\text{d}^3 \sigma_{pp \to \ell^+ N}}{\text{d} x_1 \, \text{d} x_2 \, \text{d} \hat t} &= \frac{\pi \alpha_2^2}{12 \, \hat s^2}
  \frac{\hat t \left(\hat t - m_N^2 \right)}{\left( \hat s - M_{W_R}^2 \right)^2 + \left( \Gamma M_{W_R} \right)^2}\, \times
  \\
  &  \sum_{ud} \left |V_{ud} V_{\ell N} \right |^2 \left(f_u(x_1) f_{\overline d}(x_2) + 1 \leftrightarrow 2 \right) \, ,
\end{split}
\end{equation}
where $\Gamma$ is the total decay width of $W_R$, and $\hat s$ and $\hat t$ are the partonic Mandelstam 
variables. 
Integrating over $\hat t$ and the PDFs, we get the total cross-section for $pp \to \ell N$
\begin{equation} \label{eq:SigPart}
\begin{split}
  &\sigma_{p p \to \ell^+ N} =
  \frac{\pi \alpha_2^2}{24 N_c} \int_{\frac{m_N^2}{s}}^1 \text{d} x_1 \int_{\frac{m_N^2}{x_1 s}}^1 \text{d} x_2 
  \\ 
& 
  \frac{\hat s}{\left( \hat s - M_{W_R}^2 \right)^2 + \left( \Gamma M_{W_R} \right)^2}
  \left(1 - \frac{m_N^2}{\hat s} \right)^2 \left(2 + \frac{m_N^2}{\hat s} \right) 
  \\
  &\sum_{ud} \left |V_{ud} V_{\ell N} \right |^2 \left(f_u \left(x_1 \right) f_{\overline d}
    \left(x_2 \right) + 1 \leftrightarrow 2 \right) \, . 
\end{split}  
\end{equation}

\begin{figure}
  \centerline{\includegraphics[width=\columnwidth]{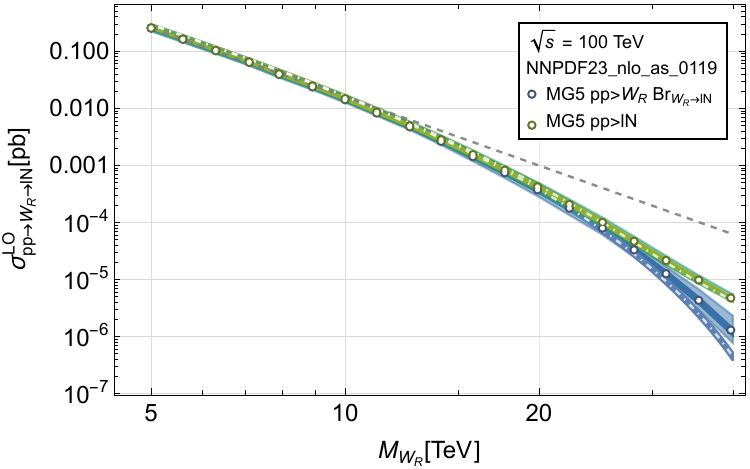}}
  \vspace*{-2ex}
  \caption{
  The total cross-section $pp \to W_R \to \ell N$, for fixed $m_N = 200 \text{ GeV}$ and $t_\beta = 0$. 
  Blue dots (solid line) represent the narrow width approximation as calculated by \texttt{MadGraph} (analytically).
  Light (dark) blue band corresponds to scale (PDF) variation.
  The green dots (solid line) represent the exact $pp \to \ell N$ result via \texttt{MadGraph} (analytics).
  Green bands also show the scale and PDF variation, which is smaller for $pp \to W_R \to \ell N$.
  The gray dashed line shows the $M_{W_R}^{-4}$ slope, normalized to 5 TeV.
  \label{figSigmaPPWR100TeV}}
\end{figure}

We compare the cross sections for $pp \to W_R$ from~\eqref{eqdSigdy} with $pp \to \ell N$ 
from~\eqref{eq:SigPart} at $\sqrt{s} = 100 \text{ TeV}$ and plot them on FIG.~\ref{figSigmaPPWR100TeV}.
These analytical calculations are shown with white dot-dashed lines and accompanied by bands that
show the uncertainties due to scale variation by a factor of 2, i.e. $\mu \in [0.5, 2] \sqrt{\hat s}$.
These are compared to the numerical results from \texttt{MadGraph}, shown with empty dots.
They are surrounded by darker bands showing the uncertainties for scale variation and the lighter 
bands for PDF member variation.
As a rule of thumb, the cross-sections fall as $M_{W_R}^{-4}$ and this na\"ive expectation is shown
in the gray dashed line.
It is a very good proxy for $W_R$ masses up to about a few 10 TeV, above which the cross-sections 
go below this simple scaling.
The narrow width approximation in blue does pretty well compared to the exact case of $2 \to 2$
scattering (shown in green) but starts to fail at about 20 TeV, missing the relevant fraction of
$W_R$ produced off-shell. It also overestimates the uncertainty in the cross-section
due to scale variation and PDF, as the uncertainty in the exact total cross section stays below
10\%.

\begin{figure}
  \centerline{\includegraphics[width=\columnwidth]{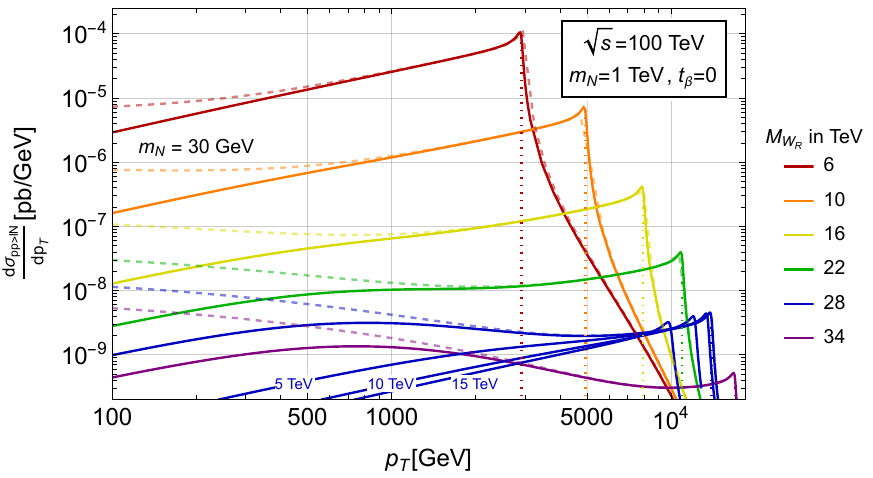}}
  \vspace{1ex}
  \centerline{\includegraphics[width=\columnwidth]{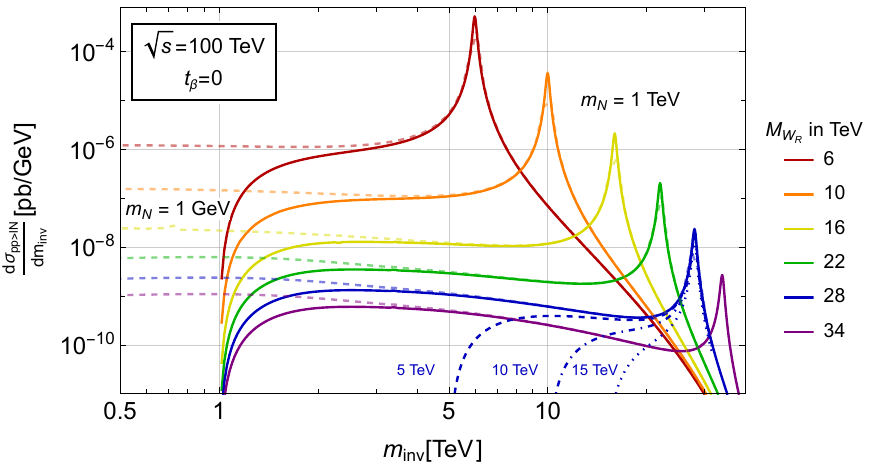}}
  \vspace*{-1ex}
  \caption{Leading lepton $p_T(\ell_1)$ distribution (top) and invariant mass of $\ell N$ (bottom)
  in the $p p \to \ell N$ production at $\sqrt s = 100 \text{ TeV}$.
  Solid lines are $m_N = 1 \text{ TeV}$ while the dotted, dot-dashed and dashed are 5, 10 and 15 TeV.
  On the $p_T$ plot we plot the vertical dotted lines of $p_T^{\max}(\ell_1)$ from~\eqref{eq:ptmax}
  and color dashed lines are for $m_N = 30 \text{ GeV}$.
  In the invariant mass plot below, the dashed lines correspond to $m_N = 1 \text{ GeV}$.
  }
  \label{fig:ptl1}
  \label{fig:invmass}
\end{figure}

To better understand the dynamics in the 100 TeV regime, we move from the total integrated cross-section 
to kinematical distributions.
Let us focus first on the $p_T(\ell) = p_T(N)$ distribution of the leading lepton, shown on the upper frame of 
FIG.~\ref{fig:ptl1} for various $M_{W_R}$ and $m_N$. 
One can distinguish the two regimes of off- and on-shell $W_R$ on the left and right portion of each line. 
Clearly, the maximal $p_T$ of the lepton (and RH neutrino) is limited by the $W_R$ mass and
by the center of mass energy via the PDFs.
For $W_R$ produced at rest, which is the relevant regime for large $M_{W_R}$, one has
\begin{equation} \label{eq:ptmax}
  p_T^{\max}(\ell_1) \simeq \frac{M_{W_R}}{2}\left(1-\frac{m_N^2}{M_{W_R}^2}\right) \, .
\end{equation}
This is derived from~\eqref{eqpTmax} with an on-shell $W_R$, i.e. by setting $\sqrt s = M_{W_R}$ and
neglecting the masses of charged leptons and protons.
As seen from the upper frame of FIG.~\ref{fig:ptl1}, where $p_T^{\max}(\ell_1)$ from~\eqref{eq:ptmax}
is plotted with vertical dotted lines, the maximal $p_T$ increases with $M_{W_R}$ when $N$ is light.
It starts to decrease when $m_N$ goes closer to the threshold of $M_{W_R}$, because $N$ is produced
progressively at rest.

\begin{figure*}
  \centerline{\includegraphics[height=.55\columnwidth]{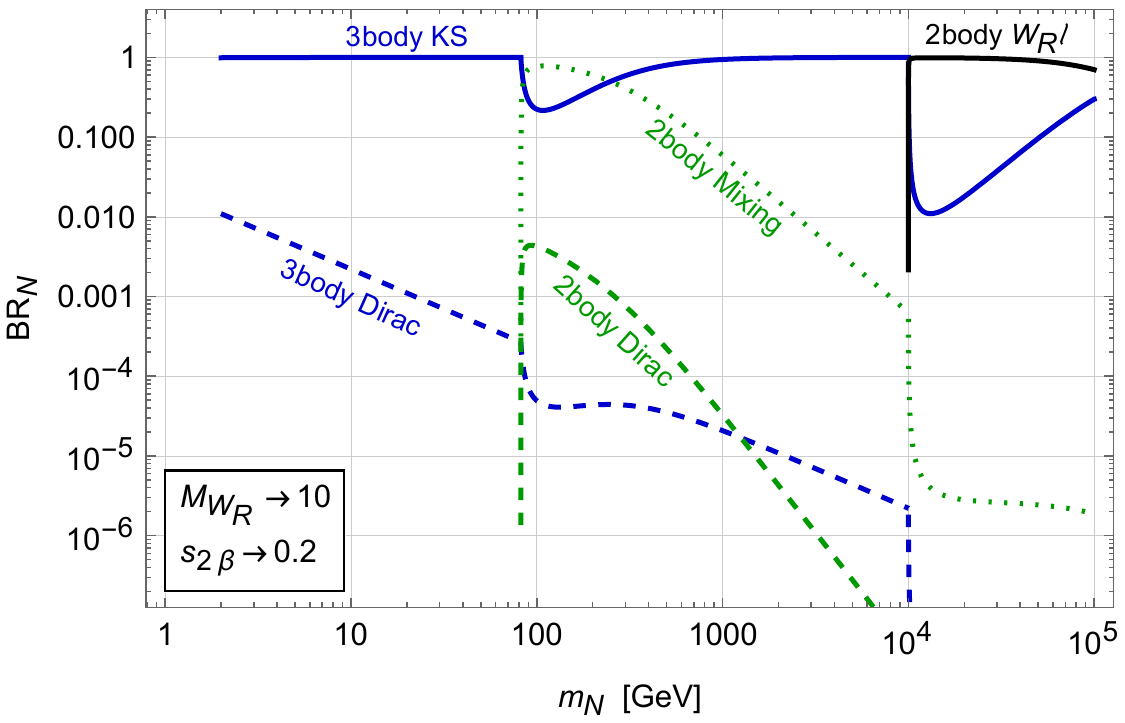}~~~~~~~~%
    \includegraphics[height=.6\columnwidth]{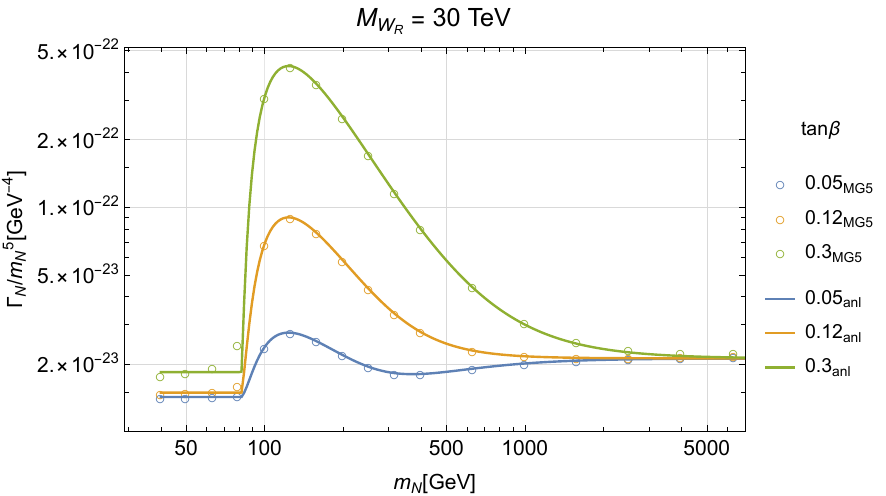}}
  \vspace*{-1ex}
  \caption{%
  Decay modes of $N$.
  On the left we show the branching ratios for $N$ decaying to different decay channels.  
  On the right, we plot the $t_\b$ dependence of the total $N$ width in the relevant mass range. 
  The empty circles refer to the numerical output from \texttt{MadGraph} and the solid lines correspond 
  to the analytic rates from~\eqref{eq:NDecaytotal2body} and~\eqref{eq:NDecaymix}.
  }
  \label{fig:NDecay}
\end{figure*}

The spectrum at low $p_T$ is dominated by the off-shell $W_R$ production, which is increasingly
important for heavier $W_R$.
The available effective $W_R$ invariant mass is limited by the center-of-mass energy via the PDFs 
to a few \TeV (dashed lines). 
The cross-section also gets suppressed as $N$ gets heavier, and together with the lowering of
$p_{T}^{\max}(\ell_1)$ as described above, one ends up with a single peak in the intermediate region
(blue solid lines on FIG.~\ref{fig:invmass}).

When $t_\b$ is turned on, additional $W_R$ decays open up, specifically into $WZ$ and $Wh$, which 
slightly reduces the branching ratio to $\ell_R N$.
On the other hand, the production of $\ell_R N$ via $W$ exchange and gauge boson mixing 
becomes possible.
This leads to an increase of events at the lower end of the $m_N$ spectrum, similar to the off-shell 
$W_R$ case and is clearly favored for $m_N < M_W$.

\smallskip

In the lower frame of FIG.~\ref{fig:invmass} we display similar useful distributions of the total 
invariant mass $m_{\text{inv}}(\ell N)$ of the $\ell - N$ system.
The most obvious feature is the characteristic peak at $M_{W_R}$.
Its behaviour at lower invariant masses is also interesting.
For larger $W_R$ masses, there is a significant off-shell plateau at lower invariant masses, see
e.g.\ the $M_{W_R} = 34 \text{ TeV}$ solid and dashed lines.
This is quite sensitive to the mass of $N$ and is essentially cut off below $m_N$, as shown
in the various blue lines for $M_{W_R} = 28 \text{ TeV}$.

\medskip

Before moving to the decay of $N$, we remark that in the present work we assume for definiteness
that other possible processes in the LRSM do not interfere with the $W_R$ and $N$ production.
In particular, given the high scales involved, one may consider the possibility that the charged components 
of the bi-doublet or triplets have a mass in the probed regime. 
Their effect, together with the $W_R$ channel, considerably complicates the signatures, due to the number 
of diverse couplings and mass scales involved. 
On the other hand, such studies will become necessary, in case a signal beyond the standard
model will be observed.  
In the literature, some of these cases were considered as benchmarks, namely the $H^\pm$ or 
$\Delta_R^{\pm\pm}$ production, see e.g.~\cite{Barenboim:1996pt, Huitu:1996su, Maalampi:2002vx, 
Roitgrund:2020cge, Mohapatra:2019qid, Dev:2015kca}. 
Also a dedicated study of $\Delta^0$ would be particularly interesting, since $\Delta^0 \to NN$ channel can
reveal the spontaneous mass origin of $m_N$ and probes lepton number violation in the Higgs 
sector~\cite{Maiezza:2015lza,Nemevsek:2016enw}.
Moreover, these channels may benefit from the large gluon-fusion production cross-section at 
$\sqrt s = 100 \text{ TeV}$.

%
%
\subsection{$N$ decay} \label{sec:Ndecay}

\noindent
The RH neutrino $N$ is typically short lived if the LR scales are in the TeV region.
It decays into a secondary charged lepton $\ell_2$ and dominantly via an off-shell $W_R$ into two 
partons, i.e.\ $N \to \ell q \bar q$.
Depending on $m_N$ and the resulting boost, the signature varies: from the lepton and two distinct jets, 
to the lepton and a single jet, to a single jet including the lepton.  
For very low $m_N$, the lifetime can be long enough such that the decay happens at a macroscopical distance 
within or even outside the detector, ending up as missing energy, see FIG.~\ref{fig:layman}.

The dominant $N$ decay width is given by
\begin{equation}
\begin{split}
  \Gamma_{N \to \ell^{\pm}q_i\bar{q_j}} &= 2\, \frac{\alpha_2^2\,m_N^5}{128\pi M_{W_R}^4}	
  \left|V^{\text{\tiny CKM}}_{ij}\right|^2 \times
  \\ 
  & \left( 1 - 8 x + 8x^2 - x^4 - 12 x^2 \log x\right) \, ,
  \label{eq:NDecaytotal2body}
\end{split}
\end{equation}
where $x = m_q^2/m_N^2$ is the heavier quark mass of the two $m_q = \max (m_i, m_j)$.
In case $m_N$ goes below $M_W$, the same final states (with opposite quark chirality) can also be obtained 
via the standard $W$ exchange and LR gauge boson mixing, by multiplying~\eqref{eq:NDecaytotal2body} 
with $\sin(2 \beta)^2$.  

In turn, two body decay channels $N \to \ell W$ open up as soon as $m_N > M_W$, both in the presence 
of LR mixing $\xi_{LR}$ or from the mixing angle $\t$ that connects the left and right handed neutrinos via the 
Dirac mass term. 
These two can be grouped together into
\begin{align} 
  \Gamma_{N \to \ell^{\pm}W^{\mp}} &= \frac{\alpha_2}{8} m_N 
  \left(\theta ^2 + \xi_{LR}^2\right) \left(\frac{1}{x} - x \right) \, ,
  \label{eq:NDecaymix} 
\end{align}
with $x = M_W^2/m_N^2 < 1$, where we dropped a small interference term.

Finally, above $m_N > M_{W_R}$ the $N \to \ell W_R$ channel becomes dominant. 
It is obtained by the above formula~\eqref{eq:NDecaymix} by replacing $M_W \to M_{W_R}$ and 
$\theta^2 +  \xi_{LR} ^2\to 1$.

\begin{figure}
  \includegraphics[width=\columnwidth]{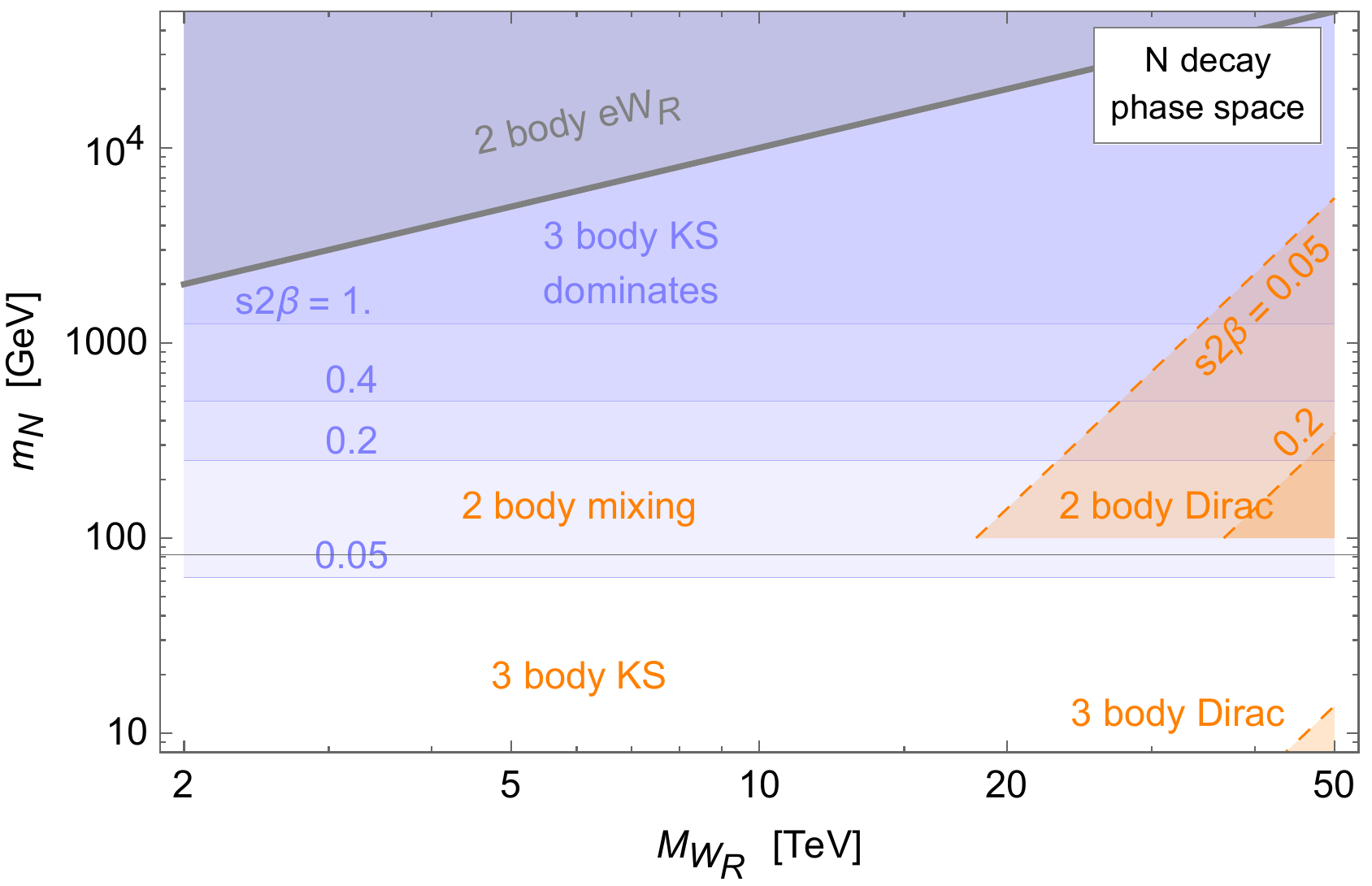}
  \vspace*{-5ex}
  \caption{Dominant $N$ decay channels as a function of $M_{W_R}$, $m_N$ and for various choices of $t_\b$. 
  The standard three body KS decay dominates everywhere for $m_N<M_{W_R}$, except for $m_N$ just above 
  $M_W$, where $N$ can decay to $W\ell$ depending on $t_\b$ via $\xi_{LR}$, and for very large $M_{W_R}$, 
  where decays via the Dirac mass can be relevant.}
  \label{fig:NDecayPhase}
\end{figure}

The relative weight of the various $N$ decay channels described above can be understood collectively
in a ``spaghetti'' plot, presented in FIG.~\ref{fig:NDecay} (left), which we exemplify for the case
of $M_{W_R} = 10 \text{ TeV}$ and a moderate value of $s_{2\b} = 0.2$.
The presence of the gauge boson mixing allows for the two-body $N$ decay, starting from $m_N \geq M_W$, 
to a few-$100$--1000\,\GeV, depending on $t_\b$.

On the right frame of FIG.~\ref{fig:NDecay} the effect of $t_\b$ on the total $N$ width can be appreciated for
$t_\b=0.05, 0.12, 0.3$.
It is evident that $t_\b$ only impacts the light $N$ below  few-100\,\GeV.
It is worth recalling here that the allowed vaues of $t_\b$ are different in case the LR symmetry is chosen to be 
either $\P$ or $\C$.
In the former case of $\P$ it is limited by flavor bounds near the value $t_\b \simeq 0.12$ at low 
$M_{W_R}$~\cite{Bertolini:2019out}, while in the latter case it is free, up to the perturbativity limit of
$t_\b \lesssim 0.5$. 
In this work we adopt $t_\b = 0.3$ as our highest benckmark value.

The dominance of the various decay channels while varying the model parameters in the
$M_{W_R}$--$m_N$ plane, can be appreciated also in FIG.~\ref{fig:NDecayPhase}. 
Here it is notable that the two-body mixing mediated decays become dominant in a region just above $M_W$, 
as will be evident in the final results.

The processes mediated by Dirac neutrino masses turn out to be subleading, or relevant only in the
hardly accessible high $M_{W_R}$ and displaced $N$ regime. 
The reason lies in the small magnitude of the Dirac masses, after recalling that they are in general not free, but 
predicted by the LRSM model in connection with the Majorana neutrino mass matrix~\cite{Nemevsek:2012iq}.

\smallskip

In fact the regime of very light $N$ is particularly interesting and promising.
For $m_N \lesssim 100 \,\GeV$ the decay can be appreciably displaced from the primary vertex (depending
also on the $W_R$ mass) while for even lower $m_N$ the decay happen most of the times outside the
detector, appearing as missing energy signature.  
These regions are naturally overlapping as we will show below.  
The displaced decay regime was studied in~\cite{Nemevsek:2018bbt} for the LHC.  
For the FCC, the prospects and sensitivities for observing displaced vertices are strongly dependent on the
inner detector design and vertexing challenges, so that, albeit very interesting, are clearly premature.
The missing energy signature on the other hand are straightforward and only weakly depend on the detector size. 
We will analayze it below.

\begin{table*}
  \def\hhh{\hphantom{\bf 0.0009523}}
  $
  \begin{array}{|c|c|c|c|c|c|c}
    \hline
  \qquad  \text{Backgrounds [pb]}
  \qquad    &\quad  \verb!w+12j!\quad  & \verb!DY+12j! & \verb!vv+012j! & \verb!tt+01j! \\
  (\sqrt s = 100\,{\rm TeV}) &  \hhh   &    \hhh    & \hhh  &     \hhh     \\ 
  \hline
  \texttt{xptj},\texttt{xptl}>50    &  5700       & 1000    &  180     & 480       \\
  \texttt{xptj},\texttt{xptl}>500   &  4.0        & 0.45    &   0.110  &  0.031    \\
  {}+ \texttt{xptl}>1000   &  0.46       & 0.030   &   0.017  &  0.0045  \\
  {}+ \texttt{misset} < 500     &  0.39   & 0.030 &   0.011  &  0.0028 \\
  \hline
  {}+ \texttt{xptj},\texttt{xptl}>1500\;\text{(detector)}& \bf 0.047 &\bf 0.0025  & \bf 0.001 &\bf 0.000012 \\
  \hline
  {}+ \text{k-factors} & \bf 0.023 &\bf 0.0017  & \bf 0.0015 &\bf 0.000024 \\
  \hline
  \end{array}
  $
  \caption{
  Table of background cross sections versus various cut requirements.
  All the cross sections are in pb and the cuts are in GeV. 
  For each case, the subprocesses with at least one final lepton were selected.
  For any jet and lepton, a minimal momentum is assumed with $\texttt{ptj} > 20$, $\texttt{ptl}>10$.
  At the analysis level we impose even stronger cuts on the leading jet/lepton with $\texttt{xptj},\texttt{xptl}>1500$.
  }
\label{tab:cutflow}
\end{table*}

%
%
\section{Backgrounds} \label{sec:Backgrounds}
\noindent Let us focus on the backgrounds for the cases when $N$ decays inside the detector (we
discuss the missing energy background later on).
Given the signature characteristics, namely at least one highly energetic prompt lepton, low missing 
energy, and one or more jets, we identify the following possible SM processes contributing as backgrounds: 
\begin{enumerate}
  \item $W$ boson plus one or two jets;
  \item Drell Yan $\ell^+ \ell^-$ plus one or two jets;
  \item diboson production $VV$ with up to two jets.
  \item $t\bar t$ plus up to one jet; 
\end{enumerate}
For all of these SM backgrounds we require the presence of at least one charged lepton.
For example, in the $VV$ samples we force at least one of the bosons to decay leptonically.

All of the backgrounds are generated using \texttt{MadGraph} 3.3.2 and 2.8.0~\cite{Alwall:2007st},
hadronized using \texttt{Pythia} 8~\cite{Sjostrand:2007gs}.
For detector simulation we used \texttt{Delphes} 3~\cite{deFavereau:2013fsa}, adopting the provisional 
FCC card~\cite{FCCcard}. 
The parton level processes are generated at tree level with jet matching. 
In addition, a review of the literature for NLO and Electroweak (EW) corrections brings the following $k$-factors: 
\begin{enumerate}
  \item for \verb!w+12j!, NLO$+$EW corrections imply~\cite{Mangano:2016jyj} a striking reduction of 
  50\% or more, especially at high $p_T \gtrsim 10 \,\TeV$ as considered here, so that we apply a 
  50\% reduction;
  \item for \verb!DY+12j!, the NLO$+\gamma$-induced corrections also bring a reduction of up to 
  50\%, so that we adopt a correction of $-30\%$; 
  \item for \verb!vv+012j!, the NLO corrections lead to an enhancement of circa 1.5; 
  \item for \verb!tt+01j!, NLO enhances by a factor of $\sim2.5$ and EW corrections bring a reduction of 
  20\%, so that an increase of 100\% is a safe estimate.
\end{enumerate}
At the same time, the signal is subject to a $k$-factor of 1.2--1.5~\cite{Mitra:2016kov} for the range of 
$W_R$ masses that we consider here.  
All of these estimates are clearly affected by their own uncertainties, motivating further studies
for increasing the precision at 100 TeV.

\begin{figure}
  \includegraphics[width=\columnwidth]{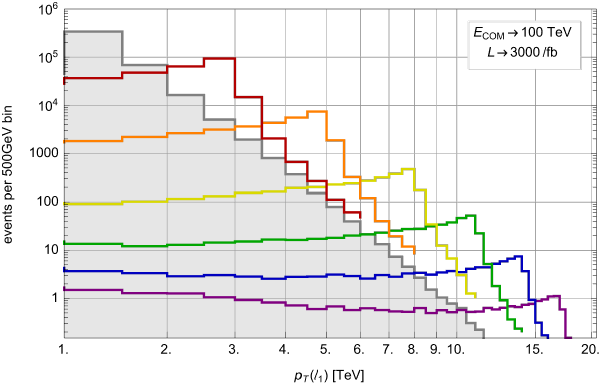}
  \vspace*{-5ex}
  \caption{
    Distribution of leading lepton $p_T$ before cuts, for the background (gray) and various signal scenarios, 
    $M_{W_R} = 6, 10, 16, 22, 28, 34 \text{ TeV}$ from left to right with a fixed $m_N = 1 \text{ TeV}$.}
  \label{fig:PTL}
\end{figure}

As discussed above, after $W_R$ decays, the prompt lepton and the subsequent leading jets
typically carry high momentum on the order of $M_{W_R}/2$. 
On the other hand, backgrounds typically concentrate at lower $p_T$s, reaching at most a few TeV. 
For illustration we plot the leading lepton $p_T$ distributions in FIG.~\ref{fig:PTL}.  
As a result, the sensitivity to the signal can be efficiently improved by restricting the leading lepton and 
jet momenta to be of the order of TeV.
At generator level we require both a leading jet and leading lepton to have $p_T > 1 \text{ TeV}$,
using the $\texttt{xptj}$ and $\texttt{xptl}$ parameters.
This reduces the background cross sections substantially, so that enough statistics can be gathered by 
Monte Carlo, even for an integrated luminosity of 3/ab. 
The $\slashed E < 500 \text{ GeV}$ cut brings in a further reduction, without significantly 
impacting the signal.
Finally, we chose to impose even stronger cuts $\texttt{xptj}$, $\texttt{xptl}>1500$ at detector level,
which further reduces the first backgrounds by $10$, and the softer \verb!tt+01j! by more 
than $10^{3}$.

The efficiency of the above cumulative cuts on the signal varies approximately from 40\% at low 
$M_{W_R}\sim 7\,$TeV to about 90\% at highest $M_{W_R}\sim 40\,$TeV (the signal is largely
off-shell and moves to softer $p_T$).
%
The cut flow and final cross sections for the background are reported in TAB.~\ref{tab:cutflow}.  
The most dominant process is \verb!w+12j!, and the other processes can be safely neglected in the
present study.

As evident from FIG.~\ref{fig:PTL}, if one wished to focus solely on the higher $W_R$ masses at  
higher luminosity, then more stringent cuts could even be imposed from the beginning.
This would further reduce the need for large samples in the generated background.
We opt instead for very minimal cuts and a larger statistics, which should reliably simulate also
the lower values of $M_{W_R}$ (still above the reach of the LHC) and lower luminosities.
To assess the sensitivity we employ a cut-free method~\cite{Nemevsek:2018bbt}, as described in 
section~\ref{sec:Sensitivity}.

%
%
\section{Signal} \label{sec:Signature}

\noindent
The signal was simulated with the same settings and cuts as the background.
We used the LRSM model file~\cite{LRSMmix-model} at LO, which was introduced in~\cite{Roitgrund:2014zka} 
and updated from~\cite{Maiezza:2015lza}.

The single prompt lepton $\ell_1$ from $W_R$ decay is typically well isolated and can serve as a
high efficiency trigger. 
At the same time, the $N$ decay products are always very energetic, as shown in the previous sections.  
This naturally happens for light $N$ ($m_N\ll M_{W_R}$), which is boosted and whose decay products  
typically have $p_T\sim M_{W_R}/3$.
Likewise, the heavier $N$ ($m_N\lesssim M_{W_R}$), features an energetic
{\em secondary} lepton and jets that have similarly a large momentum $\sim m_N/3$.

\begin{figure}
  \includegraphics[width=\columnwidth]{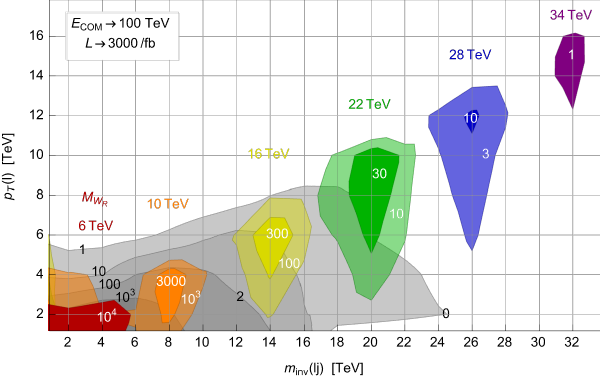}
  \vspace*{-4ex}
  \caption{%
  Distribution of events in 2$\times$2\,TeV bins in the space of leading lepton $p_T(l_1)$ versus 
  $m_{\text{inv}}(l_1j_1)$.  
  Background is shown underneath in gray, and overlayed regions correspond to signal scenarios with 
  $M_{W_R}=6,10,16,22,28,34\,$TeV left to right, and fixed $m_N=1$\,TeV (for clarity only the highest 
  contours are shown). 
  The signal moves progressively to a background free region as $M_{W_R}$ becomes heavier.}
  \label{fig:PTLMLJ}
\end{figure}

At the detector level, due to isolation limitations, it is not always possible to separate all the $N$ decay 
products, especially when they are boosted and produce a single jet that contains the secondary 
charged lepton $\ell_2$.  
Our approach is thus to reconstruct the $W_R$ invariant mass of the (leading) jet, together with one 
(leading) lepton, or possibly with two leptons if they are isolated.  
The single lepton plus jet variable $m_{\text{inv}} (l_1j_1)$ is more appropriate for the light (boosted) RH
neutrino regime, while the two lepton plus jet variable $m_{\text{inv}} (l_1l_2j_1)$ is sensitive to
the higher RH neutrino masses. 
We also always consider leptons and jets with a minimal $p_T(\ell) = 10 \text{ GeV}$ and $p_T(j) = 50 \text{ GeV}$.

In FIG.~\ref{fig:PTLMLJ} we plot the distribution of events in the $p_T(l_1)$--$m_{\text{inv}} (l_1j_1)$ plane, 
both for the background and for a selection of signal scenarios with fixed $m_N = 1 \text{ TeV}$. 
As one can see for increasing $M_{W_R}$ the signal peaks progressively outside of the background region. 
This is characteristic of $s$-channel resonance searches and is particularly promising.

\begin{figure*}
  \includegraphics[width=1.8\columnwidth]{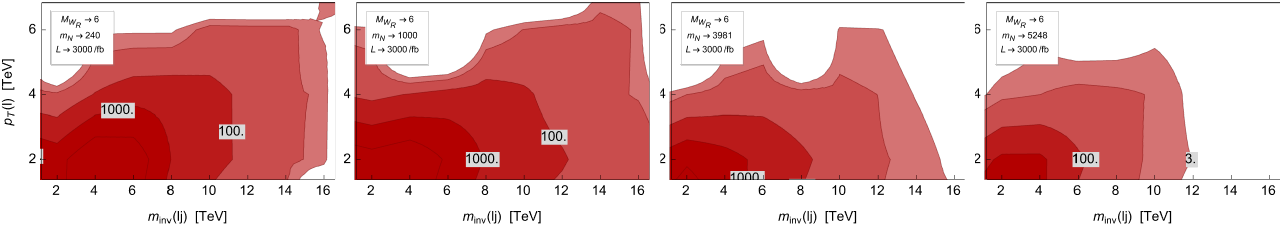}  
  \includegraphics[width=1.8\columnwidth]{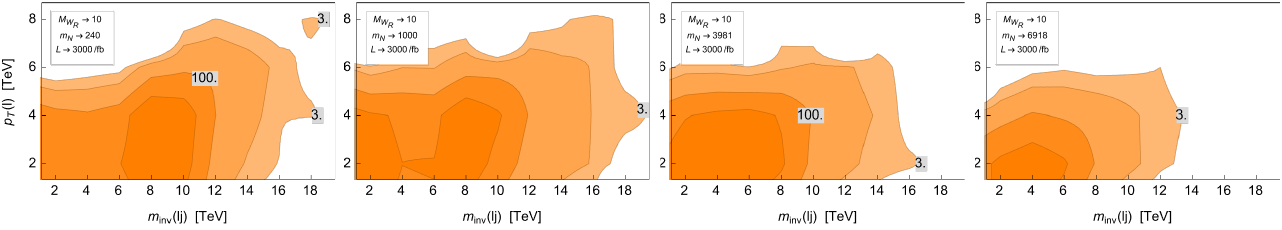}  
  \includegraphics[width=1.8\columnwidth]{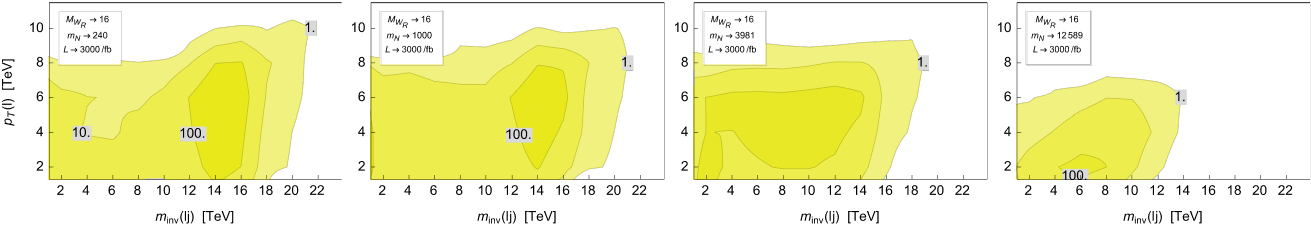}  
  \includegraphics[width=1.8\columnwidth]{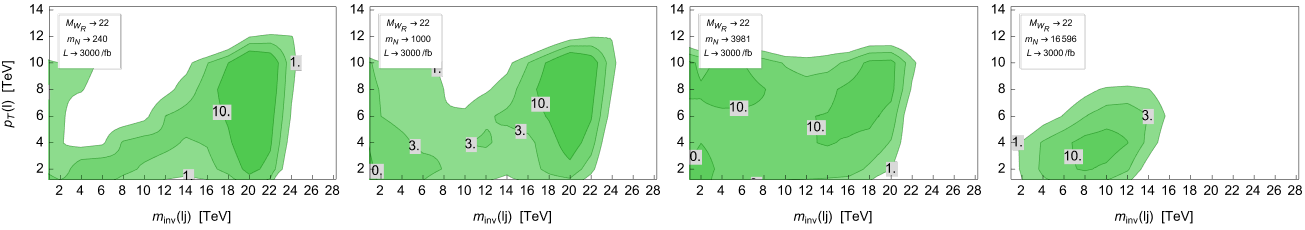}
  \includegraphics[width=1.8\columnwidth]{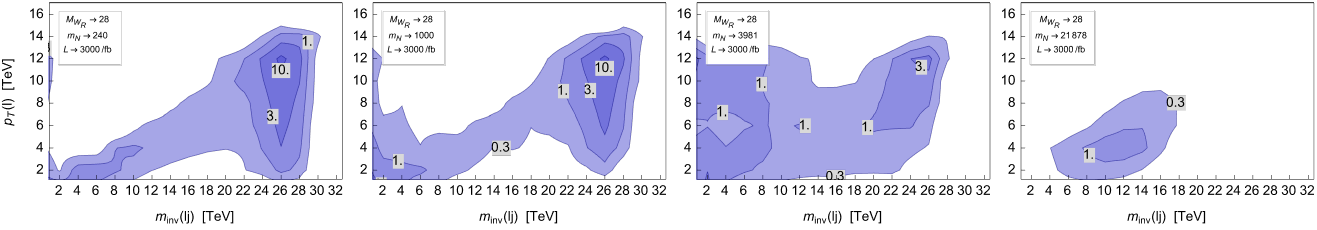}
  \includegraphics[width=1.8\columnwidth]{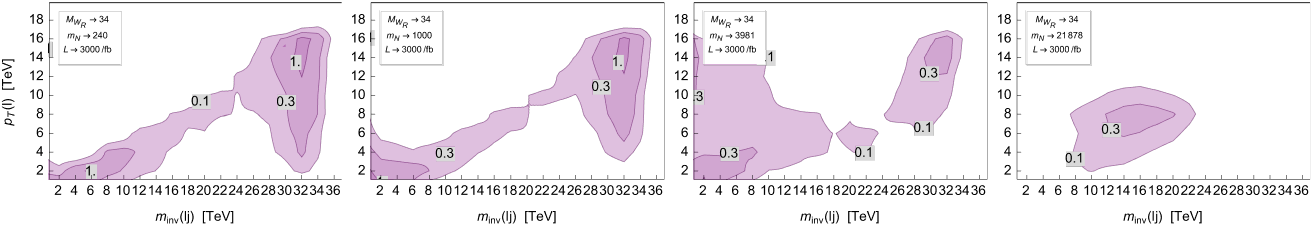}
  \caption{
  Distribution of leading lepton $p_T(l_1)$ versus $m_{\text{inv}}(l_1j_1)$, for various heavy neutrino 
  masses, and fixed $M_{W_R}$.
  }
  \label{fig:PTLMLJ_mn}
\end{figure*}

Varying the RH neutrino mass leads to various distributions that are shown in FIG.~\ref{fig:PTLMLJ_mn}.
One can observe that because the RH neutrino decays to a further lepton and jets, and
because at detector level one can not distinguish between leptons, it may happen
that the secondary lepton from $N$ decay is harder and takes the role of the first. 
This happens at large $m_N$ masses, when $p_T(\ell_2) \simeq m_N/3 > p_T^{\max}(\ell_1)$ from (\ref{eq:ptmax}). 
Solving for $m_N$, we find $m_N/M_{W_R}=(\sqrt{10}-1)/3\simeq 0.72$, with $p_T^{\max}\simeq M_{W_R}/4$. 
This applies to the rightmost frames in FIG.~\ref{fig:PTLMLJ_mn}.

Similar distribution of events appear in the $p_T(l_1)$--$m_{\text{inv}} (l_1l_2j_1)$ plane, at
large $m_N$ masses.  
In the next section we will take into account both channels and estimate the sentitivity in the 
entire $M_{W_R}$--$m_N$ plane.

\begin{table*}
\vspace*{1ex}
$
\begin{array}{|lcc|rcccccccccccc|}
\hline
  & \!\!\! \mathcal{L} = 3 \,\mathrm{ab}^{-1} \!\!\!\!\!\!\!\!\!\!\!
                                                              && M_{W_R}\!: &  10 & 10 & 10 & 10 &~~~ 18 & 18 & 18 & 18 &~~~ 32 & 32 & 32 & 32 \\
\text{variable}              & \text{range}     & \text{\# bins} & m_{N}\!: & 240 & 1000 & 3981 & 9550 &~~~ 240 & 1000 & 3981 & 16596 &~~~ 240 & 1000 & 3981 & 28840 \\
\hline                                                                                                                                                       
p_T(\ell_1)                  & \{1.5,31.5\}\,\TeV  & {16}                  &   & 106 & 78.5 & 80.8 & 2.19 &~~~~ 25.8 & 26.0 & 25.1 & 1.21 &~~~ 3.43 & 3.78 & 2.73 & 0.326 \\[0.5ex]
m_{\text{inv}}(\ell_1 j_1)       & \{1.5,41.5\}\,\TeV   & {20}               &   & 123 & 89.7 & 86.9 & 2.99 &~~~ 29.5 & 28.2 & 26.2 & 1.61 & ~~~ 4.06 & 4.28 & 3.25 & 0.531 \\[0.5ex]
m_{\text{inv}}(\ell_1 \ell_2 j_1) & \{1.5,41.5\}\,\TeV    & {20}              &   &  124 & 94.0 & 109. & 15.7 &~~~ 29.4 & 28.6 & 29.6 & 6.03 &~~~ 4.05 & 4.32 & 3.60 & 0.992 \\[0.5ex]
\hline
\end{array}
$
\caption{
  The first column reports the chosen grid binning variables, their range and number of corresponding bins. 
  The columns on the right correspond to sensitivities in $\s$ obtained with $3/\,\text{ab}^{-1}$.
  Subsequent rows show the progression/optimization of the sensitivity after adding in turn each binning variable to 
  the grid, and the bottom row represents our final sensitivity.
  The selection of points in the $m_N - M_{W_R}$ parameter space for progressively large $m_N$ ranges from the 
  single lepton to the two isolated leptons regime.}
\label{tab:sensflow}
\end{table*}

%
%
\section{Sensitivity} \label{sec:Sensitivity}

\noindent
While a fair idea of the reach may be obtained from FIG.~\ref{fig:PTL} with a sliding cut on $p_T(\ell)$ as a 
function of $M_{W_R}$, the further dependence on $m_N$ makes this procedure unfeasible.  
A simpler and optimal method to assess the sensitivity for any choice of model parameters was devised 
in~\cite{Nemevsek:2018bbt}.   
It consists of splitting the background and signal events in a multidimentional binning along a few relevant 
observables, and defining the overall sensitivity as the sum in quadrature of single bin sensitivities:
\begin{equation} \label{eq:sensitivity}
  \Sigma^2 = \sum_{i\in\text{bins}}\frac{s_i^2}{s_i+b_i} \, ,
\end{equation}
where $s_i$ and $b_i$ are the expected number of signal and background events in each bin (see
FIG.~\ref{fig:PTLMLJ} for an example of a two-variables binning of signal and background). 
The method is quite robust with respect to binning variations, with a systematic uncertainty that can 
be suitably controlled. 
We refer to~\cite{Nemevsek:2018bbt} for the illustration and the theoretical discussion of the method.

%
%
\SEC{The KS and LJ signature.}
The binning grid in this case concerns three observables: $p_T(\ell)$, $m_{\text{inv}}(\ell j)$ and
$m_{\text{inv}}(\ell \ell j)$, which are reported in the first column of TABLE~\ref{tab:sensflow}. 
Each of them spans the range (above the minimal cut) in bins of $2\,\TeV$.

As discussed above, $p_T(\ell)$ is already a strong discriminator between the signal and background.
The other two variables address the reconstruction of the $M_{W_R}$ invariant mass, and are useful
to boost the sensitivity in the low ($m_{\text{inv}}(\ell j)$) and heavy ($m_{\text{inv}}(\ell \ell j)$) RH neutrino 
mass regimes.

\begin{figure}
  \centerline{\includegraphics[width=.68\columnwidth]{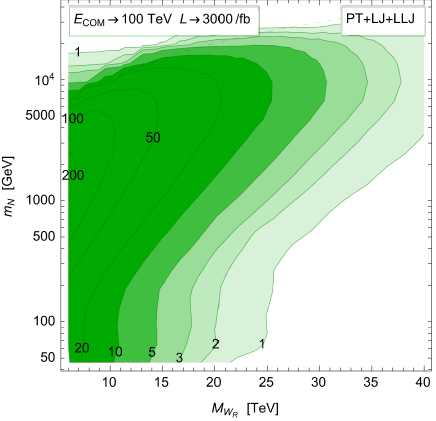}}
  \centerline{\includegraphics[width=.68\columnwidth]{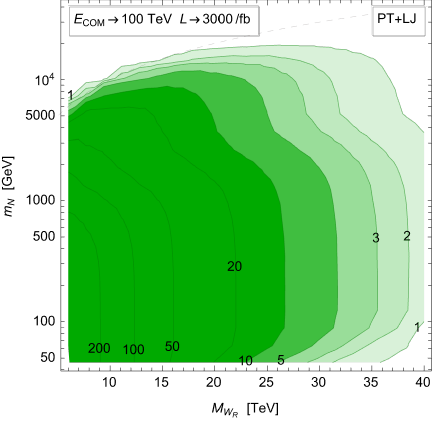}}
  \vspace*{-1ex}%
  \caption{Sensitivity (number of $\sigma$s) of the resolved two lepton LLJ (upper) and a single lepton LJ (lower) channels.}
  \label{fig:KSvsLJ}
\end{figure}

\begin{figure}
  \includegraphics[width=1\columnwidth]{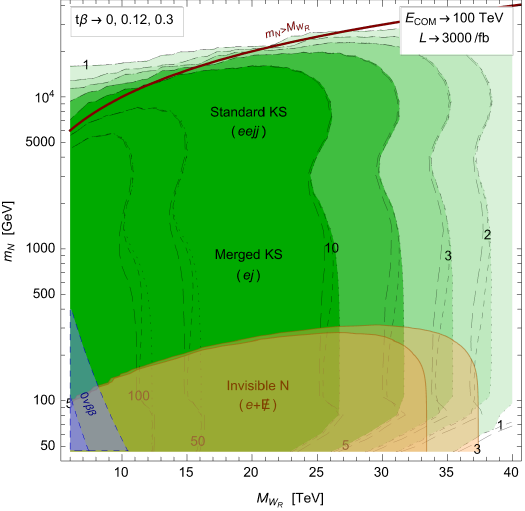}
  \vspace*{-4ex}%
  \caption{%
    The green areas show the final KS plus LJ sensitivity (in number of $\s$s) achievable with 3/ab
    integrated luminosity. 
    We show also the dependence on $t_\b = 0, 0.12, 0.3$ (dotted, dashed, long-dashed). 
    The overlayed orange shaded region in the lower part of the frame displays the 3
    and 5\,$\s$ sensitivity to the $\ell+\slashed E$ signature.}
  \label{fig:final}
\end{figure}

In TABLE~\ref{tab:sensflow} we also report the resulting sensitivity for a selection of signal points in the 
$M_{W_R}$--$m_N$ parameter space, where the successive table rows display the 
sensitivity obtained by adding in turn the corresponding variable to the binning.
One can notice the increase in sensitivity in the second line in particular for light $N$, and the
strong increase in the last line, especially for large $N$ masses.
In FIG.~\ref{fig:KSvsLJ} we display the sensitivity in the $M_{W_R}$--$m_N$ plane, for the
separate channels where the final signature has $LLJ$ (upper) or $LJ$ (lower) only. 
One can appreciate the complementarity of the two channels for low and high RH neutrino masses. 
The final combined sensitivity is shown in FIG.~\ref{fig:final} for an integrated luminosity of
$L = 3 \text{ ab}$.  It is quite notable that the combined reach of the two channels together is around
$37/35/32 \text{ TeV}$ at $2/3/5 \sigma$ of C.L..
Also for an early run with a low integrated luminosity of $L = 30 \text{ fb}$, the 5\,$\s$ discovery reach 
is around 15\,\TeV.

\medskip

At the lower end of $M_{W_R} \lesssim 15 \text{ TeV}$, the FCC-hh can probe also the heavy $m_N > M_{W_R}$
regime, where $W_R$ is produced off-shell, but with a sufficiently large invariant mass to generate an $N$ with
mass up to $m_N \simeq 15 \text{ TeV}$.

\medskip

FIG.~\ref{fig:final} also shows the impact of the presence of $t_\b$. 
For $m_N>M_W$, we see a depletion of the observed signal rate, as discussed in 
section~\ref{sec:Ndecay}. 
On the other hand, the production via $W$ opens up below $M_W$ and leads to an increase. 
However, this happens in the light $N$ regime, where $N$ decay is progressively displaced. 
Therefore, the dependence on $t_\b$ would be of major interest in future studies of displaced 
decays at FCC-hh.

\SEC{The missing energy signature.}  
For light $m_N \lesssim 300\,\GeV$, the $N$ decay length in the lab frame becomes long enough, such 
that the probability of it decaying outside of the detector becomes sizeable. 
Experimentally this shows up as a prompt lepton plus missing energy, which is the signature usually 
assumed in searches for a sequential $W' \to \ell \nu$~\cite{CMS:2022krd}.

For the FCC-hh we assume a conservative detector size of 5 meters and calculate the expected number 
of those events, where $N$ decays entirely outside of the detector, while $\ell_1$ always remains prompt.
The technical details of this analytical calculation are described at the end of Appendix~\ref{app:DY}.
To compare with the estimated expected SM backgrounds, we separate the events into bins of transverse 
mass $m_T$, as considered in~\cite{ATLAS:2017jdr} and~\cite{CMS:2022krd}, but we rescale the cross-sections to 
$\sqrt{s} = 100 \text{ TeV}$.
The background turns out to be dominated by single $W$ production and sub-dominant Drell-Yan, $t \bar t$ 
and multi-jet components.

The final result is shown as a shaded orange region and covers the lower part of FIG.~\ref{fig:final}.
It demonstrates that in this channel the $3(5)\,\s$ expected FCC reach extends up to
$M_{W_R} \simeq 37(33) \text{ TeV}$ and up to $m_N \simeq 300 \text{ GeV}$.  
The sensitivity reach here is thus slightly higher than in the KS and LJ channels, like it already happens 
at the LHC~\cite{Nemevsek:2018bbt}, but shows a nice complementarity.  
The invisible channel reach estimate would be slightly reduced in case a much larger detector size 
will be chosen, but at the advantage of the other channels discussed above. 
In addition, the possibility of detecting displaced $N$ decays, not yet considered at this stage, would 
boost the sensitivity of KS in the low $m_N$ regime, as it was argued in~\cite{Nemevsek:2018bbt}.

\medskip

An important point concerns the connection between colliders and neutrinoless double beta decay 
($0\nu\b\b$) in this light $m_N \lesssim 300 \text{ GeV}$ regime. 
There are a couple of new sources for the $0\nu\b\b$ rate present in the 
LRSM~\cite{Mohapatra:1979ia, Mohapatra:1980yp}, in addition to the standard light Majorana neutrino 
exchange.
While the standard double weak decay produces two outgoing electrons with \emph{left} chirality, in the 
LRSM new diagrams appear with \emph{two right} or \emph{one left and one right}-handed outgoing electron.
It turns out that the latter opposite chirality process may be the increasingly dominant one for heavier $M_{W_R}$. 
It is mediated by the exchange of two $W$'s, or by one $W$ plus one $W_R$. 
The latter option has a suppressed nuclear matrix element (see~\cite{Doi:1985dx,Hirsch:1996qw,Vergados:2002pv}
and~\cite{Barry:2013xxa} for details), one is left with the double $W$ exchange that can produce
opposite chirality electrons via the LR gauge boson mixing.
This contribution is thus driven by the magnitude of $t_\b$.

\begin{figure}
  \includegraphics[width=\columnwidth]{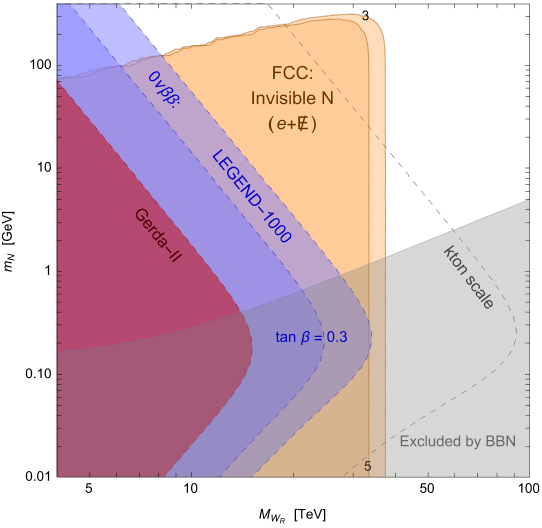}
  \caption{
    LRSM contribution to $0\nu\b\b$ from processes including the righ-right heavy $N$ exchange
    and the left-right amplitude via gauge boson mixing (for a benchmark value of $t_\b = 0.3$). 
    The red and blue shaded regions show the mass ranges saturating various $0\nu\b\b$ sensitivities, and their interplay 
    with the invisible $N$ channel at FCC in orange.
    The lower part of the diagram in shaded gray shows the exclusion from BBN.
    }
 \label{fig:legendary}
\end{figure}

%
The sensitivity to $0\nu\b\b$ experiments is shown in FIG.~\ref{fig:legendary} for a benchmark value of $t_\b = 0.3$ with
the calculable\footnote{The size of the mixing is predicted
in LRSM via the calculable Dirac mass matrix~\cite{Nemevsek:2012iq, Senjanovic:2016vxw}, but might be enhanced in 
small corners of parameter space~\cite{Barry:2013xxa}.}
seesaw Dirac mixing $U_{\nu N}\simeq(m_\nu/m_N)^{1/2}$.
We depict the region where the LRSM can saturate a possible $0\nu\b\b$ evidence, for the present and future planned 
sensitivities (GERDA-II, $m_{ee} < 92 \text{ meV}$, and LEGEND-1000, $m_{ee} < 9 - 21 \text{ meV}$). 
One can see that the region extends up to the scale of $M_{W_R} \simeq 30 \text{ TeV}$, with light $N$ in the 
(sub)\text{GeV} range. 
For $t_\b = 0.12$ the region extends up to $M_{W_R} \sim 15\,\TeV$. 
We also recall that $m_N < 0.14$--few$\,\GeV$ is excluded by the requirement that $N$ decays fast enough, in 
order not to spoil the BBN predictions~\cite{Nemevsek:2011aa, Nemevsek:2012cd}, see the gray shading. 
These come about because the $N$ with such mass is produced thermally in the early universe and then becomes
long-lived with $\tau_N \gtrsim 1 \text{ sec}$ and therefore spoils the BBN.

Finally, this region of parameter space harbors the possibility of having a (warm) dark matter(DM) candidate.
Namely, for very light masses of $N$ close to the keV scale, one can satisfy the DM abundance with 
entropy dilution~\cite{Bezrukov:2009th, Nemevsek:2012cd}.
Indeed, it has been argued in~\cite{Nemevsek:2012cd} that using the phase space suppression for the 
dilutor lifetime and the drop of relativistic degrees of freedom in $g_*(T)$ around the QCD phase transition, 
an $\mathcal O(\text{TeV})$ solution for $M_{W_R}$ may be attainable.
This requires a very light $m_N$ at about keV, in line with constraints from Dwarf spheroidals~\cite{DiPaolo:2017geq}.
Without resorting to the $g_*$ shift, a second solution for $M_{W_R}$ exists in the $20 \text{ TeV}$ range
(see FIG.~6 of~\cite{Nemevsek:2012cd}), which goes further up for heavier DM candidates (see 
also~\cite{Dror:2020jzy} for the heavy $M_{W_R}$ scenarios).

Recently, it was argued that the possibility of having a DM candidate is subject to strong constraints from the 
large scale structure data~\cite{Nemevsek:2022anh}. 
This happens because the secondary DM production from entropy injection spoils the matter power 
spectrum with potentially significant impact on the LR scale.
The fate of the DM with low $M_{W_R}$ remains to be established, but in any case the interplay with future
colliders lies precisely in connecting the DM thermal freeze-out to the missing energy signal at the FCC-hh.

Thus, on FIG.~\ref{fig:legendary} one can appreciate the interplay between $0\nu\b\b$ and the invisible 
$N$ channel at FCC: a positive $0\nu\b\b$ finding, in absence of the standard contribution (e.g.\
because of normal hierarchy, see the discussion in~\cite{Dvali:2023snt}) would imply an upper bound
on $M_{W_R}$~\cite{Nemevsek:2011aa} that we estimate in the range of $10-30 \text{ TeV}$.
This would constitute a case for looking for $W_R$ and $N$ at an FCC-hh.
Further experiments at the kton scale are envisaged to push the $0\nu\b\b$ search down to an impressive 
$\sim 1\,\meV$~\cite{Avasthi:2021lgy}.
These would connect to LRSM scales as high as $M_{W_R} \simeq 100\,\TeV$, even beyond the reach
of a 100 TeV collider.

\section{Outlook}
\label{sec:Outlook}

\noindent
While the quest for a theory of neutrino masses is still open, the LRSM stands as a unique candidate 
connecting their origin with an understanding of parity breaking in weak interactions.
Ongoing experimental efforts are at the limit of their capabilities in probing the model parameter space.
This is true for collider LHC probes, with an estimated reach of $M_{W_R} \simeq 7 \text{ TeV}$~\cite{Nemevsek:2018bbt}, 
and also low energy probes.
Most notable are current and planned $B$-meson flavour observables that will be sensitive to mass scales at 
most up to $M_{W_R} \simeq 10 \text{ TeV}$~\cite{Bertolini:2014sua}. 
It is thus important to address the prospects with the planned and proposed future experiments, with the energy 
frontier being the elected arena where to search for direct signs of new physics.
In this work we systematically estimated the reach of a hadronic ($pp$) FCC with 100\,\TeV\ center of mass energy 
in the search for a $W_R$ decaying leptonically, which has the potential to uncover lepton number violation, for 
diverse choices of $M_{W_R}$ and $m_N$, and took into account the LR gauge boson mixing.

We recalled the different signatures emerging as a function of the mass of the RH neutrino $N$: from
i) $\ell \ell j j $ in the KS process for $M_{W_R}\gtrsim m_N\gtrsim 1\,\TeV$, to ii) $\ell j_N$ in
case the $N$ decay products are merged in a single fat jet for few$\,\TeV \gtrsim m_N\gtrsim 100\,\GeV$,
to iii) the $\ell + \slashed E$ signature when $N$ decays outside the detector, for $m_N < 300\,\GeV$.  
All these channels feature at least one prompt high $p_T$ lepton, which ensures triggering and allows in reducing 
the expected SM background by many orders of magnitude.
It turns out that by requiring a minimal $p_T$ of the order of 1.5\,\TeV, the background is dominated by \verb!w+jets! 
and could be simulated to satisfactory high statistics. 
We showed indeed that, as expected, the background and signal live in different regions of kinematic observables, 
thus effectively leaving just the signal at high energies. 
As a result, the reach is mainly limited by the center of mass energy, luminosity and by quark PDFs.

We thus assessed the exclusion reach by adopting a unified binned likelihood approach~\cite{Nemevsek:2018bbt}, 
which does not require sliding windows as a function of model parameter choices. 
The results were presented in FIG.~\ref{fig:final}.  
For the KS and merged regions, i) and ii), we estimated a reach for $M_{W_R}$ as high as 35\,\TeV\ at 3\,$\s$ C.L.,
for an integrated luminosity of $3/$ab. This is similar to the reach expected for the simpler $W_R\to jj$ channel.

It is worth recalling here that within region ii), in the lightest $N$ regime of
$m_N\sim 100$--$300\,\GeV$, the RH neutrino can decay at an appreciable distance, giving rise to a
\emph{displaced jet}~\cite{Nemevsek:2018bbt}.
Its study will be very interesting as soon as definite detector geometry and sensitivities
become available.

Still, this case overlaps with the even lower $m_N$ regime (case iii) where $N$ decays outside the detector and appears 
as missing energy, thus matching with the search for $W' \to \ell \nu$.
We estimated that the $3\s$ sensitivity covers a region extending up to $M_{W_R} \simeq 37
,\TeV$ and
up to $m_N \simeq 300\,\GeV$.

This region features an interesting connection with $0\nu\b\b$ contributions from the LRSM,
especially in the light of current, planned and envisioned experiments. 
FIG.~\ref{fig:legendary} reports this interplay, showing that a possible signal at forthcoming and future $0\nu\b\b$ 
probes, in the absence of standard neutrino mass mechanism, would imply a $M_{W_R}$ below $\sim 30\,\TeV$
and $m_N$ below $\sim 100\,\GeV$, which overlaps precisely with the invisible $N$ decay at FCC here
considered.
Incidentally, cosmology presents an interesting interplay, either as a constraint from BBN or as an opportunity
for having a warm DM candidate.

\vspace*{1em}

\section*{Acknowledgments}

MN is supported by the Slovenian Research Agency under the research core funding No. P1-0035 
and in part by the research grants J1-3013, N1-0253 and J1-4389.

\appendix

\begin{widetext}
    
\section{Drell-Yan $W_R$ and $N$ production and decay rates}
\label{app:DY}

\noindent
In this appendix we review in detail the production of $W_R$ and its 
decays into the $\ell N$ final state (on shell and $2\to 2$).
%
%
\subsection{Resonance production}
\noindent
The amplitude for the production of an on-shell resonance $q(p_1)q(p_2) \to V(P)$ at parton level is 
\begin{equation}
  \mathcal A = i g \, \overline v_1 \gamma^\mu \left(v_q + a_q \gamma_5 \right) u_2 \varepsilon_\mu^{(\lambda)}.
\end{equation}
Averaging over initial spins, including the color factor and summing over polarizations $\lambda$, we have
\begin{equation}
\begin{split}
  \left| \mathcal M \right|^2 &= \frac{N_c}{4} g^2 \sum_\lambda \varepsilon_\mu^{(\lambda)} \varepsilon_\nu^{(\lambda)*} 
    \text{tr}(\gamma^\alpha \gamma^\mu \gamma^\beta \gamma^\nu \left(v_q + a_q \gamma_5 \right))  p_{1 \alpha} p_{2 \beta}
  \\
  &= \frac{N_c}{4} g^2 \left(-g_{\mu \nu} + \frac{(p_1+p_2)_\mu (p_1+p_2)_\nu}{M^2} \right)
    4 \left( p_1^\mu p_2^\nu + p_1^\nu p_2^\mu - g^{\mu \nu} p_1 p_2 \right)  \left(v_q^2 + a_q^2 \right)
  \\
  & = N_c g^2 \hat s \left(v_q^2 + a_q^2 \right),
\end{split}
\end{equation}
where $\hat s = \left(p_1 + p_2 \right)^2 = 2 p_1 p_2$. The differential cross-section is
\begin{equation}
  \text{d} \hat \sigma = (2 \pi)^4 \frac{\left| \mathcal M \right|^2}{2 \, \hat s} \text{d} \varphi,
\end{equation}
such that integration over the final state momentum gives
\begin{align}
  \hat \sigma &= \frac{\pi |\mathcal M|^2}{\hat s} \delta \left(\hat s - M^2 \right)
  = 4 N_c \pi^2 \alpha_2 \left(v_q^2 + a_q^2 \right) \delta \left( \hat s - M^2 \right).
\end{align}
For the $W_R$ cross-section, we substitute $g \to g V_{ij} /\sqrt 2 $, $v_q = a_q = 1/2$ and obtain
%
\begin{align}
  \hat \sigma_{q_i \overline{q_j} \to W_R} &= N_c \pi^2 \alpha_2 \left| V_{ij} \right|^2 \delta \left(\hat s - M_{W_R}^2 \right) .
\end{align}
%
%
Approximating the $\delta$ function with the Breit-Wigner resonance
\begin{equation}
  \pi \delta \left( \hat s - M^2 \right) \simeq \frac{\Gamma M}{\left( \hat s - M^2 \right)^2 + \left( \Gamma M \right)^2} \, ,
\end{equation}
we get the parton level cross-section
\begin{equation}
  \hat \sigma_{q_i \overline{q_j} \to W_R} = 
  N_c \pi \alpha_2 \left| V_{ij} \right|^2 \frac{\Gamma_{W_R} M_{W_R}}{
  \left( \hat s - M_{W_R}^2 \right)^2 + \left( \Gamma_{W_R} M_{W_R} \right)^2} \,.
\end{equation}

%
\medskip

\noindent{\bf Proton-proton.} To obtain the $pp$ cross-section, the partonic $\hat \sigma$ is convoluted with the PDFs. 
There is an additional $1/N_c^2$ combinatorial factor for the color connection
\begin{equation}
  \frac{\text {d}^2 \sigma_{p p \to W_R^+}}{\text{d} x_1 \text{d} x_2} = 
  \frac{\pi^2 \alpha_2}{N_c} \sum_{u,d} |V_{ud}|^2 \left(
  f_u(x_1) f_{\overline d}(x_2) + 1 \leftrightarrow 2 \right) \delta \left(\hat s - M_{W_R}^2 \right).
\end{equation}
To integrate over $\hat s$, we change integration variables from $x_{1,2}$ to the rapidity $y$ of $W_R$ and the 
partonic center of mass energy $\hat s$. 
The proton mass is small and quarks are nearly massless, therefore
\begin{equation}
  \hat s = \left( p_1 + p_2 \right)^2 = 2 p_1 p_2 = 2 x_1 x_2 P_1 P_2 = x_1 x_2 s \, .
\end{equation}
Moreover, proton beams are symmetric and energetic, such that the $W_R$ momentum is
\begin{align}
  P_W^{0,3} &= x_1 P_1^{0,3} + x_2 P_2^{0,3} = \frac{\sqrt s}{2} \left( x_1 \pm x_2 \right)  .
\end{align}
This setup gives the $W_R$ rapidity 
\begin{align}
  y \equiv \frac{1}{2} \ln \left( \frac{P_W^0 - P_W^3}{P_W^0 + P_W^3} \right) = 
  \frac{1}{2} \ln \left( \frac{x_2}{x_1} \right)  .
\end{align}
When $W_R$ is produced at rest, we have $P_W^3 = 0 = \sqrt{s}(x_1-x_2)/2$, which implies $x_1 = x_2$ and $y = 0$.
The product of $x_1$ and $x_2$ is fixed by the on-shell condition for $W_R$, which gives $P_W^{0 2} - P_W^{3 2} 
= M_{W_R}^2 = x_1 x_2 s$. 
Now the maximum value of $x_2^{\max} = 1$ corresponds to $x_1^{\min} = M_{W_R}^2/s$ and the maximal rapidity is
\begin{align}
  y_{\max}  = \ln \left( \frac{\sqrt s}{M_{W_R}} \right) \, .
\end{align}
This also corresponds to the situation when $W_R$ is maximally boosted
\begin{align}
  \gamma_R^{\max} = \frac{P_W^0}{M_{W_R}} = \frac{\sqrt s}{2} \left( x_1^{\min} + x_2^{\max} \right) = 
  \frac{1}{2} \left( \frac{\sqrt s}{M_{W_R}} + 1 \right) \simeq \frac{\sqrt s}{2 M_{W_R}} \, , 
\end{align}
where the last approximation is valid when $\sqrt s \gg M_{W_R}$.
Because quarks and protons are nearly massless, the $x_{1,2}$ have symmetric limits and $y_{\min} = - y_{\max}$.
We change variables using the Jacobian
\begin{equation}
  \text{d} x_1 \text{d} x_2 = \frac{\text{d} \hat s \, \text{d} y}{s} = x_1 x_2 \frac{\text{d} \hat s \, \text{d} y}{\hat s} \, ,
\end{equation}
and finally end up with
\begin{equation} \label{eqdSigdy}
  \frac{\text{d} \sigma}{\text{d} y} = \frac{\pi^2 \alpha_2}{N_c M_{W_R}^2} \sum_{u,d} |V_{ud}|^2 
  \left( x_1 f_u(x_1) x_2 f_{\overline d}(x_2) +  1 \leftrightarrow 2 \right).
\end{equation}
Both partonic fractions $x_{1, 2}$ are given by the collision energy and $W_R$ rapidity
$x_{1,2} = M_{W_R} e^{\pm y}/\sqrt s$.
The distributions in $y$ are symmetric at the LHC and extend to $\pm y_{\max}$.

%
\subsection{Resonance Decays}

\noindent
The on-shell $W_R$ resonance can decay in different ways. 
The dominant decay rates are
\begin{align} \label{eqGamWRqq}
  \Gamma \left(W_R \to q_R \overline {q'_R} \right) &= \frac{\alpha_2}{8} M_{W_R} \left(4 + (2 + x_t) (1 - x_t)^2 \right) ,
  \\ \label{eqGamWRellN}
  \Gamma \left(W_R \to \ell_R N \right) &= \frac{\alpha_2}{24} \left| V_{\ell N} \right|^2 M_{W_R} 
  \left(2 + x_N \right) \left(1 - x_N \right)^2,
\end{align}
%
%
where $x_f = 1 - m_f^2/M_{W_R}^2$, CKM is unitary and $V_{tb} \sim 1$.
Once we turn on the gauge boson mixing $\xi_{LR}$ in~\eqref{eqWLWRMixing}, the decay modes of $WZ$ and $Wh$, 
as well as the SM $W$ decays to $\ell_R N$, open up. 
They proceed with the following rates: 
\begin{align}
  \Gamma \left(W_R \to W Z \right) &=  \Gamma \left(W_R \to W h \right) = \frac{\alpha_2}{48} s_{2 \beta}^2 M_{W_R} \, ,
  \\
  \Gamma \left(W \to \ell_R N \right) &= \xi_{LR}^2 
  \frac{\alpha_2}{24} \left| V_{\ell N} \right|^2 M_W \left(2 + x_N \right) \left(1 - x_N \right)^2  .
\end{align}
At the same time the dominant rates in~\eqref{eqGamWRqq} and~\eqref{eqGamWRellN} get negligibly 
suppressed by $1 - \xi_{LR}^2$.

%
%
\subsection{$\ell N$ fermion pair production via resonance}
\noindent
Instead of the $2 \to 1$ resonance production above, we can consider the direct $2 \to 2$ scattering of $p p \to \ell N$
via the $W_R$ propagator. 
The scattering amplitude in the unitary gauge is given by
\begin{equation}
\begin{split}
  \mathcal A &= i \frac{g}{\sqrt 2} V_{ij} \overline u_{1R} \gamma^\mu u_{2R} \frac{- i g_{\mu \nu}}{\hat s - M^2 + i \Gamma M}
  i \frac{g}{\sqrt 2} V_{\ell N} \overline u_{3R} \gamma^\nu u_{4R}
  \\
  &= \frac{i}{2} \frac{g^2 V_{ij} V_{\ell N}}{\hat s - M^2 + i \Gamma M} \left( \overline u_{1R} \gamma^\mu u_{2R} \right)
  \left( \overline u_{3R} \gamma_\mu u_{4R} \right)  ,
\end{split}
\end{equation}
where the momentum part of the $W_R$ propagator vanishes, because we take quarks to be massless. 
Turning on the gauge boson mixing $\xi_{LR}$ does not modify the production much, the $\overline u_{1R} \gamma^\mu u_{2R}$
term goes into $\cos(\xi_{LR}) \overline u_{1R} \gamma^\mu u_{2R} + \sin(\xi_{LR}) \overline u_{1L} \gamma^\mu u_{2L}$.
After squaring the amplitude and summing over the spins, we get $\cos(\xi_{LR})^2 + \sin(\xi_{LR})^2 = 1$, because the R and L 
terms sum into the same expression, moreover the RL interference term vanishes in the limit of zero quark masses.
The spin averaged amplitude is
\begin{align} \label{eqMsq}
  \overline{ \left| \mathcal M \right|}^2 &= \frac{1}{16} N_c g^4 \frac{\left |V_{ij} V_{\ell N} \right |^2}{
  \left( \hat s - M^2 \right)^2 + \left( \Gamma M \right)^2} \text{tr}\left( \slashed p_1 \gamma^\mu \slashed p_2 \gamma^\nu P_R \right)
  \text{tr} \left( \slashed p_3 \gamma_\mu \slashed p_4 \gamma_\nu P_R \right),
  \\
  &= 4 N_c \alpha_2^2 \pi^2 \frac{\left |V_{ij} V_{\ell N} \right |^2}{
  \left( \hat s - M^2 \right)^2 + \left( \Gamma M \right)^2} \hat t \left( \hat t - m^2 \right),
\end{align}
where $\hat t = \left(p_1 - p_3 \right)^2 = \left(p_2 - p_4 \right)^2$. The partonic cross-section is then
\begin{align} \label{eqSigPart}
  \hat \sigma_{q_i \overline{q_j} \to W_R^* \to \ell N} &= \frac{1}{16 \pi \hat s^2} \int_{m^2 - \hat s}^{0} 
  \overline{ \left| \mathcal M \right|}^2 \, d \hat t
  \\
  &= \frac{1}{16 \pi \hat s^2} 4 N_c \alpha_2^2 \pi^2 \frac{\left |V_{ij} V_{\ell N} \right |^2}{
  \left( \hat s - M^2 \right)^2 + \left( \Gamma M \right)^2} \frac{1}{6} \left( \hat s - m^2 \right )^2 (2 \hat s + m^2)
  \\
   &= N_c \pi \alpha_2 \left |V_{ij} \right |^2 
  \frac{\sqrt{\hat s}}{\left( \hat s - M^2 \right)^2 + \left( \Gamma M \right)^2} \frac{\alpha_2}{24} \left| V_{\ell N} \right |^2 \sqrt{\hat s}
  \left(1 - \frac{m^2}{\hat s} \right)^2 \left(2 + \frac{m^2}{\hat s} \right).
\end{align}
In the narrow-width limit and expanding near the pole $\hat s \simeq M^2$, the cross-section becomes
\begin{align}
  \hat \sigma_{q_i \overline{q_j} \to \ell N} &= 
  N_c  \pi \alpha_2 \left |V_{ij} \right |^2 \frac{\Gamma M}{\left( \hat s - M^2 \right)^2 + \left(\Gamma M \right)^2}
  \frac{\frac{\alpha_2}{24} \left |V_{\ell N} \right |^2 M \left(1 - \frac{m^2}{M^2} \right)^2 \left(2 + \frac{m^2}{M^2} \right)}{\Gamma} 
  \\
  & \simeq N_c \pi \alpha_2 \left |V_{ij} \right |^2 \pi \delta \left( \hat s - M^2 \right) 
  \frac{\Gamma_{W_R \to \ell N}}{\Gamma} 
  = \hat \sigma_{q_i \overline{q_j} \to W_R} \text{Br}_{W_R \to \ell N} \, .
\end{align}
%

%
\medskip

\noindent{\bf Proton-proton.} The partonic cross-section in~\eqref{eqSigPart} is convoluted with the PDFs.
The differential cross-section is
\begin{align}
\begin{split}
  \frac{\text{d}^2 \sigma_{pp \to \ell^+ N}}{\text{d} x_1 \, \text{d} x_2} &= \frac{\pi \alpha_2^2}{24 N_c} 
  \frac{\hat s}{\left( \hat s - M^2 \right)^2 + \left( \Gamma M \right)^2}
  \left(1 - \frac{m^2}{\hat s} \right)^2 \left(2 + \frac{m^2}{\hat s} \right) \times
  \\
  &\times \sum_{ud} \left |V_{ud} V_{\ell N} \right |^2 \left(f_u \left(x_1 \right) f_{\overline d} \left(x_2 \right) + 1 \leftrightarrow 2 \right)  ,
\end{split}
\end{align}
and we integrate over the $x_{1,2}$ to get the total cross-section
\begin{align} \label{eqSigProton}
  \sigma_{p p \to \ell^+ N} &=
  \int_{\frac{m^2}{s}}^1 \text{d} x_1 \int_{\frac{m^2}{x_1 s}}^1 \text{d} x_2
  \frac{\text{d}^2 \sigma_{pp \to \ell^+ N}}{\text{d} x_1 \, \text{d} x_2} \, .
\end{align}
The final state kinematic variables $p_T^2$ and $\eta_\ell, \eta_N$ constrain the integration regions. 
Let
\begin{align}
  p_1 &= x_1 \frac{\sqrt{s}}{2} \left( 1, 0, 0, 1 \right)  ,  & p_3 &= \left(m_T \cosh \eta_N, 0, p_T, m_T \sinh \eta_N \right)  ,
  \\
  p_2 &= x_2 \frac{\sqrt{s}}{2} \left( 1, 0, 0, -1 \right)  , & p_4 &= \left(p_T \cosh \eta_\ell, 0, -p_T, - p_T\sinh \eta_\ell \right)  ,
\end{align}
where the transverse mass is defined by $m_T^2 = m^2 + p_T^2$.
At high momentum transfers considered here, the proton and charged lepton masses are negligible and
$p_1^2 = p_2^2 = p_4^2 = 0$.
However, we treat the heavy Majorana neutrino as massive with a mass $p_3^2 = m^2$.
From these, we get the $x_{1, 2}$ and the Mandelstam invariants, $\hat t = (p_1 - p_3)^2$ and 
$\hat s = (p_1 + p_2)^2$, to be
\begin{align}
  x_{1,2} &= \frac{1}{\sqrt s} \left( m_T e^{\pm \eta_N} + p_T e^{\mp \eta_\ell} \right) ,
  & \hat t &= - p_T \left( p_T + m_T e^{-\eta_{N \ell}} \right)  ,
  & \hat s &= x_1 x_2 s \, ,
\end{align}
where $\eta_{N \ell} = \eta_N + \eta_\ell$. 
The $p_T$ and rapidities are functions of $x_{1,2}$ ($s$ if fixed) and $\hat t$
\begin{align}
  p_T^2 &= \frac{-\hat t \left(\hat t + \hat s - m^2\right)}{\hat s}
  &
  e^{2\eta_N} &= -\frac{x_1}{x_2}\left(1+ \frac{\hat s}{\hat t} \right)
  &
  e^{2\eta_\ell} &= -\frac{x_2}{x_1}\left(1+ \frac{\hat s}{\hat t - m^2} \right)
  \\ \label{eqpTetalN}
  &= \frac{-\hat t \left(\hat t + \tau s - m^2\right)}{\tau s}\,,
  &
  &= -\frac{1}{\rho}\left(1+ \frac{\tau s}{\hat t} \right),
  &
  &= -\rho \left(1+ \frac{\tau s}{\hat t - m^2} \right),
\end{align}
where we introduced
\begin{align}
  \tau &= x_1 x_2 \, , &
  \rho &= \frac{x_2}{x_1} \, .
\end{align}
The $\tau$ variable is useful because it corresponds to the invariant mass of the $\ell N$ pair, namely
$m_{\text{inv}}^2(\ell N) = (p_3 + p_4)^2 = \hat s = x_1 x_2 s = \tau s$.
These variables simplify the imposition of cuts and efficiencies.
From the matrix element in~\eqref{eqMsq} and the definition of the cross-section in~\eqref{eqSigPart}, we have
\begin{align} \label{eqdsigdx12dth}
  \frac{\text{d}^3 \sigma_{pp \to \ell^+ N}}{\text{d} x_1 \, \text{d} x_2 \, \text{d} \hat t} &= \frac{\pi \alpha_2^2}{12 \, \hat s^2}
  \frac{\hat t \left(\hat t - m^2 \right)}{\left( \hat s - M^2 \right)^2 + \left( \Gamma M \right)^2}
  \sum_{ud} \left |V_{ud} V_{\ell N} \right |^2 \left(f_u(x_1) f_{\overline d}(x_2) + 1 \leftrightarrow 2 \right)  .
\end{align}

\medskip

\noindent{\bf Invariant mass.} 
The $m_{\text{inv}} \in [m, \sqrt{s}]$.
Taking into account that $m_{\text{inv}}^2 = \hat s = x_1 x_2 s = \tau s$, we get that
\begin{align} \label{eqdsigdx12dminv}
  \frac{\text{d} \sigma_{pp \to \ell^+ N}}{\text{d} m_{\text{inv}}} &= 
  \frac{\text{d} \sigma_{pp \to \ell^+ N}}{\text{d} m_{\text{inv}}^2} \frac{\text{d} m_{\text{inv}}^2}{\text{d} m_{\text{inv}}} 
  =
  \frac{1}{s}\frac{\text{d} \sigma_{pp \to \ell^+ N}}{\text{d} \tau} 2 m_{\text{inv}}
  = \frac{2 m_{\text{inv}}}{s} \int_{m_{\text{inv}}^2/s}^1 \frac{\text{d} x_1}{x_1} \frac{\text{d}^2 \sigma_{pp \to \ell^+ N}}{\text{d} x_1 \text{d} x_2} 
  \\
  \begin{split}
    &=\frac{\pi \alpha_2^2}{12 N_c s}
  \frac{m_{\text{inv}}^3}{\left( m_{\text{inv}}^2 - M^2 \right)^2 + \left( \Gamma M \right)^2} 
  \left(1 - \frac{m^2}{m_{\text{inv}}^2} \right)^2 \left(2 + \frac{m^2}{m_{\text{inv}}^2} \right)\times
  \\
  &\times\sum_{ud} \left |V_{ud} V_{\ell N} \right |^2 \int_{m_{\text{inv}}^2/s}^1 \frac{\text{d} x_1}{x_1} \left(f_u \left(x_1 \right) 
  f_{\overline d} \left(x_2 \right) + 1 \leftrightarrow 2 \right) ,
  \end{split}
\end{align}
where $x_2 = m_{\text{inv}}^2/(x_1 s)$ and the Jacobian from $(x_1,x_2) \to (x_1, \tau)$ is equal to $1/x_1$.

\medskip

\noindent{\bf Transverse momentum.}
Along the same lines, the $p_T$  distribution is obtained by the chain rule 
\begin{align} \label{eqdsigdx12dpT}
  \frac{\text{d} \sigma_{pp \to \ell^+ N}}{\text{d} p_T} &= \int_{x_{1,2}}
  \frac{\text{d}^3 \sigma_{pp \to \ell^+ N}}{\text{d} x_1 \, \text{d} x_2 \, \text{d} \hat t} \frac{\text{d} \hat t}{\text{d} p_T} \, ,
  \\
  \label{eqdtdpT}
    \frac{\text{d} \hat t}{\text{d} p_T} &= \frac{2 p_T \hat s}{\sqrt{(\hat s - m^2)^2 - 4 p_T^2 \hat s}} \, ,
  \\ \label{eqththm2}
  \hat t \left(\hat t - m^2 \right) &= \frac{\hat s}{2} 
  \left(\hat s - m^2 - 2 p_T^2 + \sqrt{(\hat s - m^2)^2 - 4 p_T^2 \hat s} \right)  .
\end{align}
This gives us the distribution over $p_T$, shown in Fig.~\ref{fig:ptl1}. 
For $\hat t \in \mathbb{R}$, the argument of the square root in~\eqref{eqththm2} needs to be 
positive, which leads to an upper bound on $p_T$
\begin{align} \label{eqpTmax}
  p_T \in \left[ 0, \frac{s - m^2}{2 \sqrt s } \right] .
\end{align}
Furthermore, at a fixed value of $p_T$, the lower bound for $\tau = x_1 x_2$ is given by
\begin{align}
  \tau_0 = \frac{m^2 + 2 p_T^2 + 2 \sqrt{p_T^2 (m^2 + p_T^2)}}{s} \xrightarrow{p_T \to 0} \frac{m^2}{s} \, ,
\end{align}
and the integration limits in~\eqref{eqdsigdx12dpT} are given by
\begin{align}
  \int_{x_{1,2}} = \int_{\tau_0}^1 \text{d} x_1 \int_{\tau_0/x_1}^1 \text{d} x_2 \, .
\end{align}
%

%

\medskip

\noindent{\bf Invariant mass vs $p_T$.} 
Likewise, we can get the double differential distribution over $p_T$ and $m_{\text{inv}}$ by combining the two chain
rules and integrating over $x_1$
\begin{align}
  \frac{\text{d}^2 \sigma_{pp \to \ell^+ N}}{\text{d} p_T \, \text{d} m_{\text{inv}}} &= 
  \frac{2 m_{\text{inv}}}{s} \int_{m_{\text{inv}}^2/s}^1 \frac{\text{d} x_1}{x_1} \, 
  \frac{\text{d}^3 \sigma_{pp \to \ell^+ N}}{\text{d} x_1 \, \text{d} x_2 \, \text{d} \hat t} 
  \, \frac{\text{d} \hat t}{\text{d} p_T} \, ,
\end{align}
using~\eqref{eqdsigdx12dth},~\eqref{eqdtdpT} and~\eqref{eqththm2}, where again $\hat s = m_\text{inv}^2$
and $x_2 = m_\text{inv}^2/(x_1 s)$.
From~\eqref{eqdtdpT} we see that 
\begin{align}
  p_T &\in \left[ 0, \frac{m_{\text{inv}}^2 - m^2}{2 m_{\text{inv}}} \right]  ,
  \qquad\text{ for a fixed }\quad
  m_{\text{inv}} \in \left[ m, s \right]  .
\end{align}
The contours of the double differential cross-section are interesting in that they are dominated by a narrow diagonal line relative to off-shell $W_R$ production, and a some broader regions with lower $p_T$ but fixed $m_{\text{inv}}\simeq M_{W_R}$, corresponding to on-shell $W_R$.  


\begin{description} 
\item[No cuts] Without cuts, the lower bound on $\tau$ is simply $\tau_0 = m^2/s$, as needed to produce a massive $N$. 
However, the integration limits for $\rho$ are split into two regions
\begin{equation}
  \tau \in [\tau_0, \rho], \text{ when } \rho \in [\tau_0, 1] \text{ and } \tau \in [\tau_0, 1/\rho] \text{ for } \rho \in [1, 1/\tau_0] \, .
\end{equation}
Meanwhile, $\hat t \in [m^2 - \tau s, 0]$.

  \item[$p_T$ cut] $p_T > p_{Tc}$ can be implemented within the integration limits. 
  Note that $\eta_{\ell N}$ does not depend on $p_T$. 
  From the $p_T$ equation in~\eqref{eqpTetalN}, we have
\begin{equation}
  \hat t_\pm(\tau) = \frac{s}{2} \left( \tau_0 - \tau \right) \left( 1 \pm \sqrt{ 1 - \frac{p_{Tc}^2}{s} \frac{4 \tau}{(\tau_0 - \tau)^2} } \right) ,
\end{equation}
and from the positivity of the square root, a $\rho$-independent constant lower bound
\begin{equation} \label{eqTauMpT}
  \tau_- = \tau_0 + \frac{2 p_{Tc}^2}{s} \left( 1 + \sqrt{1 + \frac{m^2}{p_{Tc}^2}} \right) \xrightarrow{p_{Tc} \to 0} \tau_0 \, ,
\end{equation}
appears. The $\tau-\rho$ integration plane is $\tau \in [\tau_-, \rho]$, when $\rho \in [\tau_-,1]$ and $\tau \in [\tau_-, 1/\rho]$ for $\rho \in [1, 1/\tau_-]$.
%
%
\item[$\eta_\ell$ cut] restriction $\left| \eta_\ell \right| >  \eta_{\ell c}$ makes the $\tau_-$ limit $\rho$ dependent. Notice that $\hat t_{\min} = 0$, so setting $\hat t = 0$
\begin{equation}
  \tau_- (\rho) = \tau_0 \left( 1 + \frac{e^{-2 \eta_{\ell c}}}{\rho}\right) \xrightarrow{\eta_{\ell c} \to \infty} \tau_0,
\end{equation}
while the $\rho$ interval comes from $\tau_-(\rho_-) = \rho_-$, and $\tau_-(1/\rho_+) = 1/\rho_+$, such that
\begin{equation}
  \rho \in [\rho_-, \rho_+] = \left[ \frac{\tau_0}{2} \left(1 + \sqrt{1 + \frac{4 e^{-2 \eta_{\ell c}}}{\tau_0}} \right), 
  \frac{1}{\tau_0} - e^{-2 \eta_{\ell c}}\right].
\end{equation}
Finally, the bound on $\hat t$ coming from~\eqref{eqpTetalN} is
\begin{equation}
  \hat t (\tau) = s \left(\tau_0  - \frac{\tau}{1 + e^{-2 \eta_{\ell c}}/\rho} \right), \quad 
  - \hat t_{\max} = - \hat t(\rho=1) = s \left( \tau_0 - \frac{1}{1 + e^{-2 \eta_{\ell c}}}\right).
\end{equation}
%
%
\item[$p_T$ and $\eta_\ell$] With both cuts acting simultaneously, the integration limits become more complex. The upper bound on $\tau(\rho)$ remains the same, however the lower bound $\tau_-$ depends on both, $p_T$ and $\eta_\ell$ cuts. More importantly, the $\rho$ interval changes as well as the limits on $\hat t$.

Let us start with the lower bound on $\tau$. Solving the quadratic equation for $\hat t$ in~\eqref{eqpTetalN} and plugging into $\eta_\ell$, we have
\begin{equation} \label{eqTauMpTeta}
  \tau_-^\pm(\rho) = \left(m^2 + p_{Tc}^2 \right) \left( 1 + \frac{e^{\pm 2 \eta_{\ell c}}}{\rho} \right) + p_{Tc}^2 \left( 1 + e^{\mp2 \eta_{\ell c}} \rho \right).
\end{equation}
The $+(-)$ applies to regions of $\rho$ above (below) 1, where the upper bound is $\rho(1/\rho)$, such that
\begin{align}
  \rho^b_+ = - \frac{e^{\eta_{\ell c}}}{2 p_{Tc}^2} &\left( e^{\eta_{\ell c}} \left(m^2 + 2 p_{Tc}^2 \right) + \sqrt{e^{2 \eta_{\ell c}} m^4 + 4 p_{Tc}^2 s} \right),
  \\
  \rho^b_- = - \frac{e^{-\eta_{\ell c}}}{2 (p_{Tc}^2 e^{2 \eta_{\ell c}} - s)} &\left( 
  e^{\eta_{\ell c}} \left(m^2 + 2 p_{Tc}^2 \right) + \sqrt{e^{2 \eta_{\ell c}} m^4 + 4 \left(p_{Tc}^2 + m^2 \right) s} \right).
\end{align}
The lower bound in~\eqref{eqTauMpTeta} should not go below the $\rho$ independent one in~\eqref{eqTauMpT}, which happens at
\begin{align}
  \rho^c_{\pm} = e^{\pm \eta_{\ell c}} \sqrt{1 + \frac{m^2}{p_{Tc}^2}}.
\end{align}
Notice that above a limiting $p_T$ cut
\begin{equation} \label{eqpTlim}
  p_{Tc}^{2 \text{ lim}} = e^{-2 \eta_{\ell c}} s - m^2,
\end{equation}
the $\eta_\ell$ cut becomes ineffective (we are back to the $p_T$ cut case above) and~\eqref{eqpTlim} 
implies an upper limit on $\eta_{\ell c}^{\text{lim}} = - \log (m/\sqrt{s})$.

\end{description}

\bigskip

\noindent{\bf $N$ as missing energy.} 
Let us compute the number of events $n_{\ell \slashed E}$, when $N$ decays outside of the detector of size 
$l_0 \sim 5 \text{ m}$. 
Events are distributed by an exponential distribution
\begin{align}
  \frac{\text{d} n}{\text{d} l} &= n_0 \frac{\exp(-l/L)}{L} \, , &
  L &= \gamma_N \beta_N \tau_N = \frac{p_T}{m \, \Gamma_N} \sqrt{1 + \left(1 + \frac{m^2}{p_T^2} \right) \sinh \eta_N^2} \, ,
\end{align}
and the total number of events is obtained by integrating from $\ell_0$ to $\infty$
\begin{align}
  n_{\ell \slashed E} &= \int_{l_0}^\infty \frac{\text{d} n}{\text{d} l} \text{d} l = \mathcal L  
  \iiint  \frac{\text{d} \rho \text{d} \tau \text{d} \hat t}{2 \rho} \varepsilon_\ell(p_T, \eta_\ell)
  \left(\frac{\text{d}^3 \sigma_{pp \to \ell N}}{\text{d} \tau \, \text{d} \rho \, \text{d} \hat t} \right) e^{-l_0/L} \, ,
\end{align}
where $\mathcal L$ is the total luminosity and $\varepsilon_\ell$ is the charged lepton selection efficiency.
Efficiencies may also include more stringent $p_T$ and $\eta_\ell$ cuts.

  \end{widetext}






\bibliographystyle{utphysmod}

\bibliography{lr-fccbib_v01}

  
\end{document}